\documentclass{JFM-FLM_Au}
\usepackage{graphicx}
\usepackage{epstopdf,epsfig}
\usepackage{newtxtext}
\usepackage{newtxmath}
\usepackage{hyperref}
\hypersetup{
    colorlinks = true,
    urlcolor   = blue,
    citecolor  = blue,
    linkcolor  = blue,
}
\usepackage[justification=justified]{caption}
\usepackage{amsmath}
\usepackage{enumerate}

\lefttitle{G. Chen, L. Gan. and P. H. Gaskell}
\righttitle{Journal of Fluid Mechanics}

\title{Effect of an aligned current on the stability of oscillatory incompressible flow past a circular cylinder}

\author{Geng Chen\aff{1}, Lian Gan\aff{1} \and Philip H. Gaskell\aff{1}}

\affiliation{\aff{1}Department of Engineering, Durham University, DH1 4LE, UK
}

\corresau{Geng Chen, \email{geng.chen@durham.ac.uk}}
\usepackage{color}
\usepackage{tikz}
\definecolor{mplmagenta}{rgb}{0.75, 0.0, 0.75}
\newcommand{\blueline}{\raisebox{2pt}{\tikz{\draw[-,blue,solid,line width = 0.4pt](0,0) -- (3mm,0);}}}

\DeclareRobustCommand{\magentaline}{\raisebox{2pt}{\tikz{\draw[-,mplmagenta,solid,line width = 0.4pt](0,0) -- (3mm,0);}}}

\newif\ifshowcomments
\showcommentstrue    % ← uncomment for draft

\definecolor{green}{RGB}{0, 139, 139} 
\definecolor{red}{RGB}{180, 6, 6}
\begin{document}

\maketitle

\begin{abstract} 
The stability of incompressible flow past a circular cylinder, subjected to a collinear superposition of steady and oscillatory forcing, is investigated within a two-dimensional Floquet framework. The flow is parameterised by the Keulegan–Carpenter number $KC \in [4,12]$, the steady-to-oscillatory velocity ratio $m \in [0,1]$, and the oscillatory Reynolds number $\Rey_m \in [20, 100]$. 
The loci of the leading Floquet multipliers, and hence flow case specific bifurcation modes, are examined by progressively reducing $\Rey_m$ to subcritical values for a prescribed $m$. 
The presence of a steady current, when $m>0.5$, 
gives rise to a period-doubling subharmonic bifurcation 
that does not occur in purely oscillatory flow, where only synchronous and quasi-periodic modes arise. For the case $\Rey_m=100$ three key features are discernable.
The first is that the neutral stability curve 
in $(KC,m)$ parameter space is strongly non-monotonic in $m$, separating regions of intrinsic stability from those characterised by single unstable modes. A Sub-region of striking mode re-stabilisation is also present beyond $m\approx0.9$, where the flow recovers a $Z_2$-symmetric state at peak Reynolds number $\approx190$, despite 
the fact that 
the steady and oscillatory components are individually unstable. Secondly, a distinct parameter regime is found to support the coexistence of two unstable modes of different types. Third, complementary direct numerical simulations show that, for cases with a single unstable mode, the linear Floquet analysis successfully predicts the saturated nonlinear state even when $\Rey_m=100$ substantially exceeds the critical Reynolds number. In contrast, for cases with mode coexistence, the quasi-periodic attractor tends to dominate the fully developed nonlinear dynamics.
\end{abstract}


\section{Introduction}\label{sec:intro}

The flow
of a steady current or oscillatory flow past a circular cylinder has been a topic of interest for decades. An important practical application featuring a combination of the two 
is relevant to wave-current met-ocean conditions that are 
encountered in marine and offshore engineering applications
\citep{Sumer2006} --
albeit at a Reynolds number several orders of magnitude larger than considered in the present work --  
the focus being almost exclusively on hydrodynamic loading and the force coefficients associated with cylindrical structures \citep{ZHOU2000,Konstantinidis2017} at low to moderate Reynolds number from $40$ to  $5200$,
or cases where the oscillatory component is applied as a weak perturbation to a dominant steady current \citep{Konstantinidis2003,Lu2011}. At a more fundamental, curiosity-driven level, the stability of both flow types has received considerable attention, as described below.

\begin{figure}
    \centering
    \includegraphics[width=0.9\linewidth]{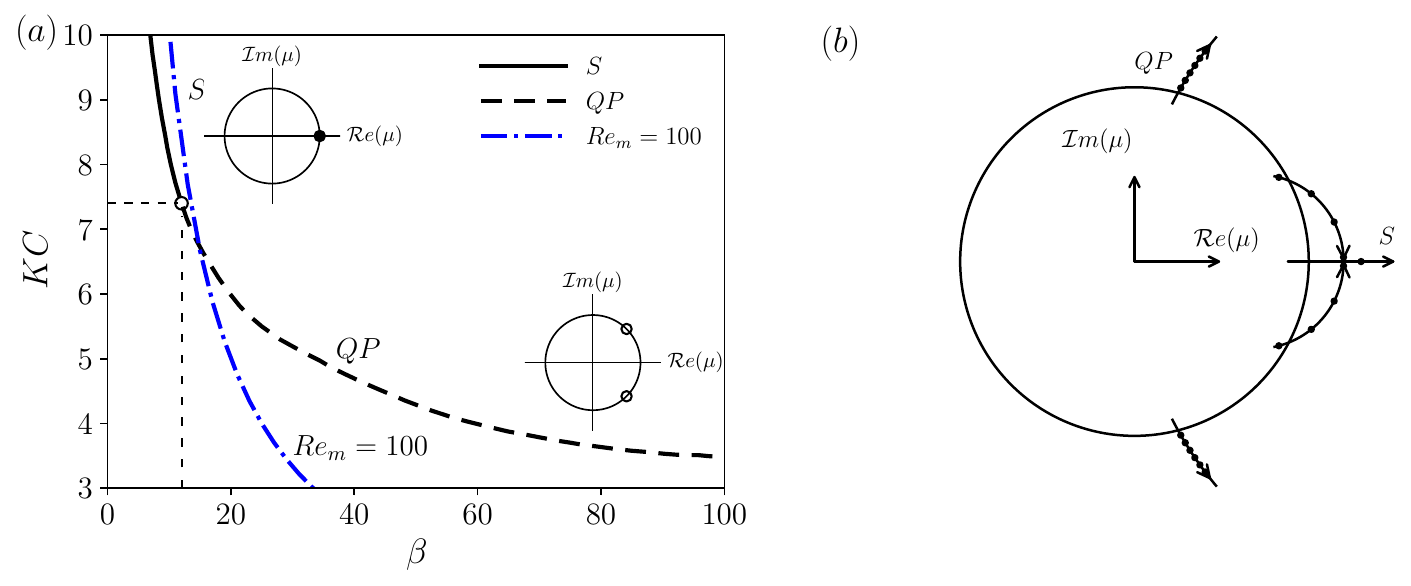}
    \caption{Purely oscillatory flow. (\textit{a}) Stability map in $(\beta, KC)$ space, indicating the onset of synchronous ($S$, solid line) and quasi-periodic ($QP$, dashed line) instability. 
    The blue dot-dashed line signifies $\Rey_m=100$. (\textit{b}) Possible Floquet multiplier $\mu$ loci in the complex plane with increasing $\Rey_m$. Arrows depict the two characteristic pathways identified in \citet{Elston2006}: \emph{S}, where a complex conjugate pair exits the unit circle and subsequently coalesce on the positive real axis, becoming real; \emph{QP}, where a complex conjugate pair exits the unit circle and remains complex. Figure modified from \citet{Elston2006}.
    } 
    \label{fig:elston_reproduce}
\end{figure}

In the case of purely oscillatory, incompressible flow (kinematic viscosity $\nu$) past a circular cylinder of diameter $D$,
bifurcation of the 
base flow is controlled by two dimensionless parameters: the Reynolds number $\Rey_m = U_m D/\nu$ and the Keulegan--Carpenter number $KC = U_m T/D$, where $U_m$ is the amplitude of the simple harmonic oscillation over the time period $T$.
Their ratio $\beta = \Rey_m/KC$ is the Stokes number. Large $\beta$ corresponds to inertia-dominated flow, while for larger $KC$ the flow excursion per cycle exceeds the diameter of the cylinder and nonlinear convection dominates \citep{Sarpkaya1984}. From their experiments, \citet{Tatsuno1990} found, depending on the combination of parameters $(KC,\beta)$,  
that the symmetry of the fundamental flow about the oscillatory axis can be broken, leading to distinctive asymmetric patterns. These observations were subsequently explained by \citet{Elston2004,Elston2006} using Floquet analysis, showing that within a two-dimensional (2D) framework only two types of primary symmetry-breaking modes arise: a synchronous mode and a quasi-periodic one. The former occurs at lower $\beta$ and is associated with a real Floquet multiplier exiting the unit circle on the positive real axis in the complex plane, i.e., a pitchfork bifurcation of the Poincar\'e map. The latter dominates at higher $\beta$ and arises from a complex-conjugate pair of multipliers that cross the unit circle, corresponding to a Neimark--Sacker bifurcation. For the range of $(KC,\beta)$ examined, no period-doubling mode was reported over the entire Poincar\'e map, with a real multiplier exiting the unit circle on the negative real axis. According to \citet{Swift1984}, as well as \citet{Marques2004} who generalised the problem to systems with additional spatial symmetries, period-doubling of a $Z_2$-symmetric limit cycle is inhibited. Consequently, within this $Z_2$ symmetry-constrained setting,  synchronous and quasi-periodic modes exhaust the generic codimension-1 bifurcation routes available to the oscillatory base flow. The corresponding neutral stability curves 
and the root loci are reproduced in figure~\ref{fig:elston_reproduce} (\textit{a}) and (\textit{b}), respectively.

For an incompressible steady current (
velocity $U_{c}$) the mechanism leading to instability is fundamentally different, it having been well-established that the Reynolds number, $Re = U_{c}D/\nu$, is the primary governing parameter.
Stability is lost at $\Rey \approx 46$ through a Hopf bifurcation, in which the reflective symmetry of the wake flow is broken into a spatio-temporal symmetry, giving rise to the canonical K\'arm\'an vortex street \citep{Mathis1987,Jackson1987,BLACKBURN_MARQUES_LOPEZ_2005,Sipp2007}. 
On increasing $Re$ further, the flow subsequently undergoes a secondary bifurcation at \(\Rey \approx 190\), leading to the development of a three-dimensional (3D) span-wise mode \citep{Barkley1996,Blackburn2003}.

The following contribution is a valuable addition to the findings mentioned above. It comprises a systematic investigation of how purely oscillatory flow past a cylinder is affected when aligned with a steady current, parametrised by the velocity ratio $m = U_c/U_m(>0)$. The work considers an oscillatory-dominant regime in which flow reversal and cycle-to-cycle vortex interactions are significant, one that has received little attention and whose fundamental linear stability characteristics remain largely unexplored. The focus is the stability characteristics of 2D flow, governed by the three parameters $KC$, $Re_m$ and $m$, laying the groundwork for a future study of 3D secondary instabilities building on the 2D limit cycles identified here --  a route known to be relevant to both steady current \citep{Barkley1996} and purely oscillatory \citep{Elston2006} flow.

The remainder of the paper is organised as follows. \S\ref{sec:methods} describes the numerical framework that underpins the fully nonlinear computations performed and Floquet stability analysis. This is followed in \S\ref{sec:result} by the results obtained beginning, when $\Rey_m = 100$, with the generation of a global stability map and a focus on the stable flow regime in    \S\ref{sec:globalmap} and \S\ref{sec:baseflow}, respectively. This is followed by the elucidation of 
leading Floquet multiplier pathways reflecting each unstable mode, together with the corresponding fully developed nonlinear flow states in \S\ref{sec:routes}-\S\ref{sec:coexist}, with the key findings regarding mode transition and interrelation covered in \S\ref{sec:Discussion}. Concluding remarks are provided in \S\ref{sec:conclusion}.

\section{Problem Formulation and Methods of Solution } 
\label{sec:methods}
\subsection{Computational Framework}
\label{sec:num_methods}

\begin{figure}
    \centering
    \includegraphics[width=0.9\linewidth]{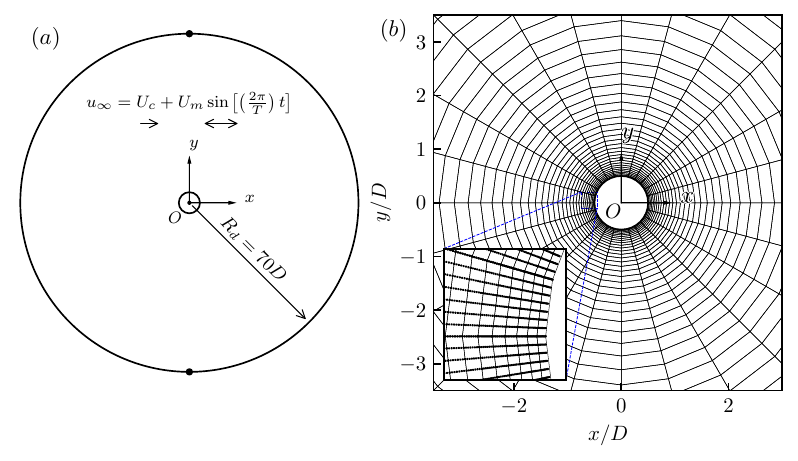}
    \caption{($a$) Schematic of the 2D circular computational solution domain, of total radius $R_d = 70D$ (not to scale), with the time-dependent free-stream velocity $u_\infty(t)$ imposed at the outer boundary. $(b)$ Domain discretisation showing the structured quadrilateral mesh employed, together with an insert showing an exploded-view in the vicinity of the cylinder surface and revealing also the high-order internal nodal distribution within individual mesh elements.}
    \label{fig:sketch}
\end{figure}

Incompressible 2D oscillatory flow, velocity $U_m$, superposed with an aligned steady current, velocity $U_c$, past a cylinder of {\color{black}{diameter}} $D$, 
is investigated numerically; the outer circular computational domain utilized has a radius $R_d = 70D$ that extends from the centre of the cylinder, 
 see figure~\ref{fig:sketch}($a$). The resulting blockage ratio is $1.4\%$, well below the $20\%$ threshold at which blockage effects have an impact on solution precision  \citep{Anagnostopoulos2004}. An O-type mesh is employed around the cylinder, providing good orthogonality at the cylinder surface, where a standard no-slip condition is enforced. 
The dimensionless velocity components applied at the outer circular boundary are: 
\begin{eqnarray}
    \frac{u_\infty(t)}{U_m} =m + \sin\left(\frac{2\pi t} {T}\right), \qquad \frac{v_\infty(t)}{U_m} = 0, 
    \label{eqn:BCU}
\end{eqnarray}
combined with the imposition of a high-order Neumann pressure condition \citep{Karniadakis1991,Blackburn2004}. 
Computations were initialised with the fluid at rest and a uniform pressure field; consequently, no attempt was made to address possible hysteresis effects.

To characterise this forcing, two Reynolds numbers are employed -- both based on the cylinder diameter $D$ and kinematic viscosity $\nu$ but differing in terms of their reference velocity -- $Re_m$ defined above and:
\begin{equation}
     \Rey_p = \frac{(U_m + U_c)\,D}{\nu} = (1 + m)\,\Rey_m
 \end{equation}
henceforth 
referred to as
peak instantaneous Reynolds number,
because $U_m + U_c$ is the largest instantaneous velocity attained during the forcing cycle. Throughout the paper $\Rey_m$ is used as the control parameter, with $\Rey_p$ quoted where comparison with the steady-flow literature is informative.

The parameters governing the combined flow are $m \in [0, 1]$ and $KC \in [4, 12]$, consistent with the investigations of \citet{Elston2006} and \citet{Ren2019}; subsequent particular flow cases are denoted by $(KC,m)$, with variation of $KC$ achieved by changing the forcing period $T$.
The direct numerical simulations (DNS) presented are for $Re_m=100$, in accordance with which the instantaneous Reynolds number within an oscillatory cycle
increases with $m$ from $\Rey_p=100$ ($m=0$) to $\Rey_p=200$ ($m=1$). 

The justification for a 2D analysis rests with earlier related investigations. \cite{Barkley1996} and \cite{Blackburn2003} showed, for a steady current, that the 2D periodic wake flow loses stability to 3D secondary modes at $Re$ 
around $190$.
In the case of a purely oscillatory flow ($m=0$) the onset of 3D symmetry breaking 
occurs only for $\Rey_m \gtrsim 200$ when $KC \lesssim 7$ \citep{Tatsuno1990,Elston2004,Tong2014};
for higher values of $KC$, although 3D flow is observed, its effect remains relatively insignificant \citep{Ren2019}.  
If the current and oscillatory components are treated as linearly decoupled for the 2D flow of interest, keeping $\Rey_p \le 200$ should, in the majority of cases investigated, be a sufficiently good approximation yielding results 
consistent with similar stability analyses \citep{Tong2014,Ren2019}. Nevertheless, it is important to recognize that their coupled effect might trigger the onset of 3D instability at a lower Reynolds number; 
determination of this threshold is beyond the scope of the present study and is left as a topic for future investigation.

The equations describing the 2D base and nonlinear saturated flow, written in dimensionless form consistent with (\ref{eqn:BCU}) -- velocity $\mathbf{u}=(u,v)$, spatial coordinates ($x, y$), pressure, $p$, and time, $t$, normalised by of $U_{m}$, $D$, $\rho U_{m}^{2}$ and $D/U_{m}$, respectively, where $\rho$ is the fluid density -- are:
\begin{subequations}\label{eqn:NS}
\begin{gather}
    \nabla \cdot \mathbf{u} =0, \\ 
    \frac{\partial \mathbf{u}}{\partial t}+(\mathbf{u} \cdot \nabla) \mathbf{u}=-
    \nabla p+\frac{1}{\Rey_{m}} \nabla^2 \mathbf{u}. 
\end{gather}
\end{subequations}
The above system was solved numerically using a spectral/\textit{hp} element method implemented in Nektar++ \citep{Cantwell2015}. The spatial discretisation employed 1415 $hp$-type spectral elements with a polynomial expansion of order $N_p=6$, enabling $p$-refinement within each element. The mesh was locally refined close to the cylinder surface, having 24 quadrilateral elements around its circumference and a minimum radial element thickness of $0.03D$ there, 
see figure~\ref{fig:sketch}($b$).  

Two far-field boundary conditions were considered. \citet{Elston2006} applied a uniform time-varying Dirichlet velocity together with a high-order Neumann pressure condition on all open boundaries, which was also adopted by \citet{Xiong2018b,Xiong2018a,Ren2019,Ju2020}. An alternative approach \citep{Tong2017,An2011,Tang2018} prescribes the same Dirichlet velocity at the inlet but specifies the streamwise pressure, or pressure gradient, at the outlet explicitly through the momentum equation with a zero-gradient velocity. As noted by \citet{Xiong2018a}, provided that the computational domain is sufficiently large ($>40D$), both approaches result in a negligible difference between the numerical results obtained. In the present work, the first approach was followed.  

Time integration was performed using a second-order velocity-correction scheme with a fixed time step, $\Delta t$, ensuring satisfaction of the Courant-Friedrichs-Lewy (CFL) condition  $\left| U \right|\Delta t/\Delta l<0.5$ throughout the simulation, where
$|U|$ and $\Delta l$ are the local velocity magnitude and element length scale, respectively. 
A systematic mesh convergence and validation study concerning the numerical procedure was performed for a purely oscillatory flow ($m=0$), and found to compare very favourably with corresponding data from the literature; details are provided in Appendix~\ref{appA}.

The streamwise and transverse force coefficients $C_d$ and $C_l$ are defined as:
\begin{equation}
    C_d(t) = \frac{2F_x(t)} {\rho U_{\mathrm{ref}}^{2} D}  \hspace{20pt} 
    C_l(t) = \frac{2F_y(t)} {\rho U_{\mathrm{ref}}^{2} D},
    \label{eqn:cdcl}
\end{equation} 
where $F_x$ and $F_y$ are the corresponding fluid forces in the $x$ and $y$ direction, respectively; see figure~\ref{fig:sketch}($a$). The temporal development of $C_l$ and its spectrum are used to characterise the global flow features, e.g. whether the fully developed flow is synchronous, subharmonic or quasi-periodic.  The reference velocity $U_{\mathrm{ref}}$ for the combined flow can be defined in different meaningful ways. 
Here $U_{\mathrm{ref}} = \sqrt{U_m^2 + U_c^2}=U_m\sqrt{\left(1 + m^2\right)}$ is adopted to account for the contribution of both components. 

\subsection{Floquet Analysis}\label{sec:floquetM}
Floquet stability analysis requires a base flow $\left[\mathbf{U}(x,y,t),P(x,y,t)\right]$ that is strictly $T$-periodic, i.e. $\mathbf{U}(x,y,t)=\mathbf{U}(x,y,t+T)$, $P(x,y,t)=P(x,y,t+T)$, which bifurcates at a critical Reynolds number
via breaking the reflective symmetry about $y=0$. Following \citet{Elston2006}, the base flow is obtained by enforcing the relevant symmetry at every instant:
\begin{equation}
    K : \left(x,\,y,\,U,\,V,\,p,\,t\right) \;\mapsto\; \left(x, \,-y, \,U, -V, \,p, \,t\right),
    \label{eqn:baseSym}
\end{equation}
where $(U,V)$ are the two components of $\mathbf{U}$. This is achieved by imposing a symmetry boundary condition on $y = 0$ and solving equations~(\ref{eqn:NS}) for the upper half-domain ($y \ge 0$), with the same outer-boundary and cylinder surface conditions described in \S\ref{sec:num_methods}. The full-domain base flow is constructed by enforcing the reflective symmetry in (\ref{eqn:baseSym}). The latter was computed for at least $200T$ to allow transients to decay, after which 32 equally spaced snapshots were extracted over the final period. Fourier-interpolation in time then provides $[\mathbf{U},P]$ at any instant required by the Floquet analysis.

The temporal growth of small-amplitude perturbations to the velocity and pressure fields, $\mathbf{u^{\prime}}(x, y, t)$ and $p^{\prime}(x, y, t)$, is governed by the linearised Navier–Stokes equations; 
\begin{subequations}
\label{eq:perteq}
\begin{gather}
\nabla \cdot \mathbf{u^{\prime}} =0,
\label{eq:cont_pert} \\
 \frac{\partial \mathbf{u^{\prime}}}{\partial t}=-\mathbf{U} \cdot \nabla \mathbf{u^{\prime}} -\mathbf{u^{\prime}} \cdot \nabla \mathbf{U}-\frac{1}{\rho}\nabla p^{\prime}+\frac{1}{\Rey_m} \nabla^2 \mathbf{u^{\prime}},
\label{eq:mom_pert}
\end{gather}
\end{subequations}
and solved over the full domain. The perturbation satisfies the Dirichlet condition $\mathbf{u^{\prime}}= 0$ on the cylinder and on the entire outer boundary, and the perturbation pressure satisfies the high-order Neumann condition of \citet{Karniadakis1991} on these same boundaries.  
In compact form, the system (\ref{eq:perteq}) can be written as:
\begin{equation}
    \frac{\partial\mathbf{q'}}{\partial t}=\mathcal{L}(t)\mathbf{q'},
    \label{eq:comp}
\end{equation}
where $\mathbf{q'}=[\mathbf{u'},p']$ and $\mathcal{L}(t+T)=\mathcal{L}(t)$. In the Floquet framework \citep{iooss2012elementary}, the solution of (\ref{eq:comp}) can be written as:
\begin{equation}
    \mathbf{q'}(t+T)=\mathbf{M}(t)\mathbf{q'}(t),
\end{equation}
where the $T$-periodic operator $\mathbf{M}(t)$ is the monodromy matrix referenced to arbitrary time $t$ (or phase due to periodicity), and often defined with $t=0$. Eigenvalues of $\mathbf{M}(t)$, $\mu = \exp(\sigma T)$, are the Floquet multipliers, with $\sigma$ being the (complex) Floquet exponent, and are phase independent. Consequently, the velocity perturbation can be expressed as the sum of the Floquet modes $\hat{\mathbf{u}}(t) = \tilde{\mathbf{u}}(t)\exp(\sigma t)$, with $\hat{\mathbf{u}}(t+T) = \mu\hat{\mathbf{u}}(t)$, where $\tilde{\mathbf{u}}$ is the $T$-periodic Floquet eigenfunction for velocity, $\tilde{\mathbf{u}}(t+T)=\tilde{\mathbf{u}}(t)$, which is the associated eigenfunction of the system (\ref{eq:comp}). A Floquet mode grows exponentially in time when the magnitude of its associated multiplier satisfies $|\mu| > 1$, while all perturbations decay and the base flow is linearly stable when $|\mu| < 1$ for every multiplier. Marginal stability corresponds to $|\mu| = 1$, where the perturbation returns to the same amplitude after each period of the base flow.

\begin{figure}
\centering
\includegraphics[width=1\linewidth]{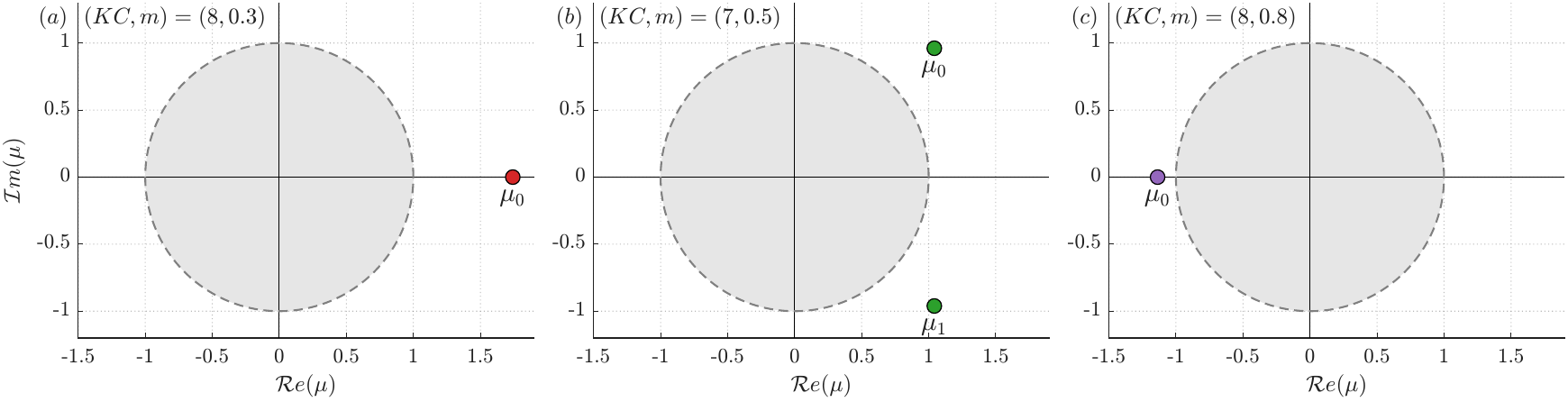}
\caption{Leading Floquet multipliers ($\mu_0$, $\mu_1$) in the complex plane for the three representative unstable Floquet mode types evaluated when $Re_m = 100$: $(a)$ synchronous; $(b)$ quasi-periodic; $(c)$ subharmonic, or period-doubling. 
}
\label{fig:floquet_subplots_grid}
\end{figure}

Owing to the complex nature of $\mu$, its argument, $\arg(\mu)$, i.e. the phase of the multiplier, characterises the time periodicity of the bifurcated flow, and this can be visualised in the complex plane. The unstable leading Floquet mode can evolve  as one of three types: 
\begin{enumerate}
    \item[(i)] synchronously ($S$) with $T-$periodic base flow if $\arg(\mu)=0$: $\mathcal{R}e(\mu)>1$, $\mathcal{I}m(\mu) =0$, i.e., a real $\mu$ located on the positive real axis outside the unit circle, figure \ref{fig:floquet_subplots_grid} ($a$).
    \item[(ii)] quasi-periodically ($QP$), if $0<\arg(\mu)<\pi$, i.e., a complex-conjugate pair with $\mathcal{I}m(\mu) \ne 0$, figure \ref{fig:floquet_subplots_grid} ($b$).

    \item[(iii)] subharmonically ($SH$) at $2T$ if $\arg(\mu)=\pi$: $\mathcal{R}e(\mu)<-1$, $\mathcal{I}m(\mu) =0$, i.e., a real $\mu$ located on the negative real axis outside the unit circle, figure~\ref{fig:floquet_subplots_grid} ($c$).
\end{enumerate}
Taking into account the symmetry about the real axis, for the results that follow it is only necessary to present the upper half-plane $\mathcal{I}m(\mu) \geq 0$.

At a $\Rey_m$ 
higher than the critical value, more than one multipliers may lie outside the unit circle in the upper half-plane, and with 4 possible combinations of the three mode types listed above: $S+S$, $S+QP$, $SH+SH$  or $QP+QP$ (for some cases at $Re_m<100$);
$S+SH$ and $SH+QP$ are not observed.   
Such states are referred to as modal coexistence. However, they do not necessarily exit the unit circle at the same $Re_m$, i.e., not necessarily codimension-2.

The Floquet stability map, in $(KC,m)$ space, presented in \S\ref{sec:globalmap}, is constructed for 
fixed $Re_m=100$. The leading eigenvalues $\mu$ of $\mathbf{M}(t)$ are computed using the matrix-free Arnoldi method implemented on ARPACK. In this approach, each action of $\mathbf{M}(t)$ on a perturbation field is evaluated by time integrating~\eqref{eq:perteq} over one period. The numerical specifications comprise 
a Krylov subspace dimension $\kappa = 32$, an iterative tolerance $10^{-6}$, a maximum of 2000 Arnoldi iterations, and 10 requested converged eigenvalues per solve. The implementation within Nektar++ follows the approaches of \citet{Barkley1996} and \citet{ Elston2004, Elston2006}. 

Validation against the neutral stability boundary for the $m=0$ cases determined by \citet{Elston2006} is presented in Appendix~\ref{sec:FloquetValid}; the agreement obtained supports the reliability of the present Floquet framework and numerical underpinnings. The sensitivity of the eigenvalue computation to boundary conditions and domain shape was also evaluated, confirming that the computational settings adopted in this study leads to negligible differences ($<1\%$) across different boundary conditions and domain shapes. 

To determine the loci of Floquet multipliers in the complex plane for different bifurcation modes, $\Rey_m$ is progressively reduced from $\Rey_m=100$ to values as low as $20$ for representative cases. It is also worth mentioning that the locations of the multipliers shown in the complex plane, such as in figure~\ref{fig:floquet_subplots_grid}, are exact for the given $(KC,m)$;  the same is true for all figures of this type presented subsequently. 

\section{Results and discussion} \label{sec:result}

\subsection{Global stability map when  $Re_m=100$} \label{sec:globalmap}

\begin{figure}
\centering
\includegraphics[width=1\linewidth]{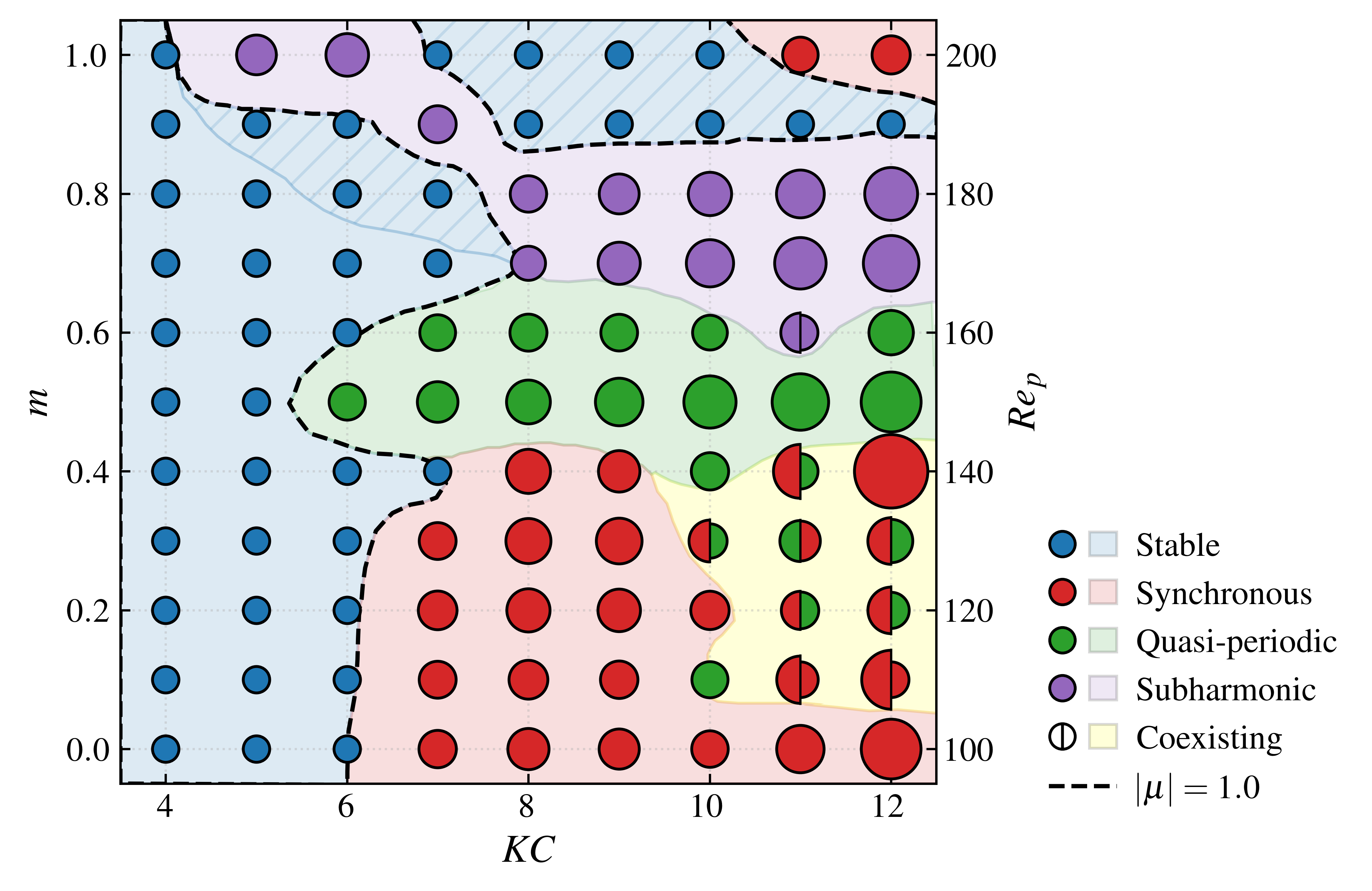}
\caption{Floquet stability map in the $(KC, m)$ space at $\Rey_m = 100$. The right vertical axis is $\Rey_p = \Rey_m(1+m)$. Marker colours indicate the stability state of the flow: stable (blue), synchronous ($S$, red), quasi-periodic ($QP$, green), and subharmonic ($SH$, purple). Split markers denote two Floquet multipliers located outside the unit circle, with left and right halves representing the leading and sub-leading multipliers (see figure~\ref{fig:floquet_coexist2} for examples). Marker size is proportional to the magnitude of the leading multiplier $|\mu_0|$ for unstable cases. The dashed black lines are  neutral stability curves ($|\mu_0| = 1$) determined by linear interpolation of $|\mu_0|$ for neighbouring cases. 
Distinct stability classification regimes are also distinguished by the pale, background colouring indicated in the legend. Re-stabilisation sub-regions are identified with a hatched overlay,
where the flow at lower $Re_m$ can be unstable.
}
\label{fig:stability_map}
\end{figure}

For the problem of interest, the nature of the flow is depicted in terms of its dependence on $(KC \in [4,12], m \in [0,1])$, when $Re_m=100$, as a colour-map in figure~\ref{fig:stability_map}. Four flow types arise: stable, synchronous ($S$), quasi-periodic ($QP$), and subharmonic ($SH$). The stable mode corresponds to the magnitude of the leading Floquet multiplier $|\mu_0|\leq1$; the three unstable modes ($|\mu_0|>1$) are classified according to $\arg(\mu_0)$, as explained in \S\ref{sec:floquetM}, with their corresponding marker size scaled proportionally with $|\mu_0|$. 
Certain cases, for large $KC$, have two multipliers located outside of the unit circle in the upper half-plane (a complex-conjugate pair counted as one), having the same or two different classifications; they are identified as split markers, with left and right halves indicating the leading, $\mu_0$, and sub-leading, $\mu_1$, multipliers, respectively -- each is coloured according to its classification and scaled by its magnitude. Over the $(KC,m)$ range investigated, no cases were found  exhibiting three or more multipliers located outside of the unit circle.

Figure~\ref{fig:stability_map} reveals a rather complex stability landscape. For purely oscillatory flows ($m=0$, bottom row), the dependence of the flow stability on $KC$ agrees with the findings of \cite{Elston2006}, figure~\ref{fig:elston_reproduce}(\textit{a}). That is,  symmetry breaking occurs at $KC\approx7$, leading to a synchronous unstable mode. The map also shows that for $KC\lesssim6$ the superposed $U_c$ component does not modify the flow stability characteristics in general, except when $m>0.9$, where a subharmonic unstable mode emerges (pale-purple zone).  No $S$ or $QP$ unstable mode is observed in this $KC$ range. 

For $7\lesssim KC\lesssim10$,  the flow is unstable, the mode of which changes from $S$ (pale-red zone) to $QP$ (pale-green zone), followed by $SH$ (pale-purple zone) as $m$ increases to roughly 0.9.  This $SH$ mode is inaccessible for $m=0$ due to the shift-reflect $Z_2$ constraint of a purely oscillatory flow \citep{Swift1984,Marques2004}.  
In general, the magnitude of the leading multiplier $|\mu_0|$ increases with $KC$, indicated by the size of the marker. The pattern is clearer for the $QP$ and $SH$  modes in the interval $0.4< m<0.9$ than for the $S$ mode; $|\mu_0|$ does not show strong $m$ dependence for a given $KC$. 

A striking and perhaps counter-intuitive observation is that a small further increment in $m$ leads to flow re-stabilisation (the hatched pale-blue shaded region at  the upper-right of figure~\ref{fig:stability_map}). That is, reflective symmetry is re-established, notwithstanding the fact that it occurs at the highest Reynolds number range examined, $Re_p\gtrsim190$, where purely oscillatory and purely steady flow  are individually 2D unstable. The mechanism behind this 
re-stablisation of the flow 
is elaborated upon in \S\ref{sec:island}. There is also an interesting case when $KC=10$ and $m=0.1$, a green marker surrounded by red ones, which is discussed in \S\ref{sec:coexist}.   

\begin{figure}
\centering
\includegraphics[width=1\linewidth]{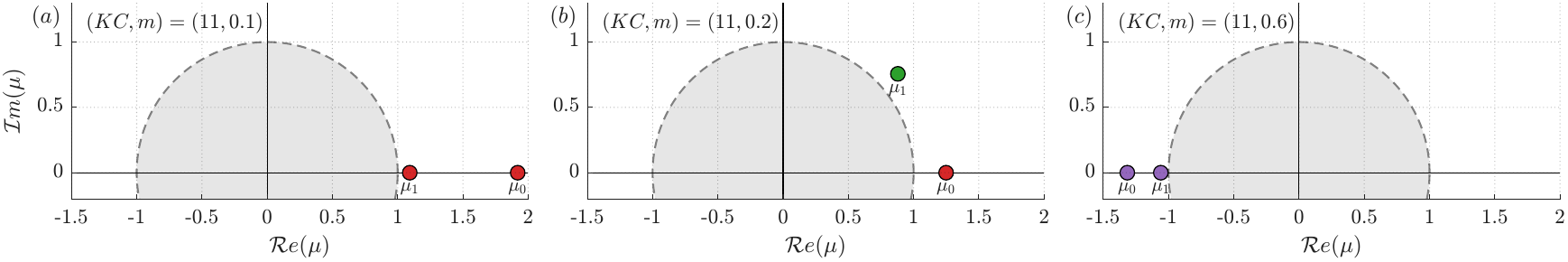}
\caption{The three possible types of modal coexistence at $KC = 11$, which demonstrate the rich dependence on $m$ of the instability mode selection. ($a$) $m = 0.1$: $S+S$; ($b$) $m = 0.2$: $S+QP$; ($c$) $m = 0.6$: $SH+SH$. }
\label{fig:floquet_coexist2}
\end{figure}

The range $KC>10$ maintains the $m$-dependent stability pattern, 
other than the flow becoming synchronously unstable again as $m$ increases from 0.9 to 1, viz. $(KC,m)=(11,1)$ and $(12,1)$. The associated mechanism responsible 
is not discussed, since at $Re_p=200$, 3D effects are likely to be influencing the flow that are not captured by the 2D model. Mode coexistence is found within the range $0.1\lesssim m\lesssim0.5$, denoted by the split markers in figure~\ref{fig:stability_map}, the three possible types of which are illustrated in figure~\ref{fig:floquet_coexist2}. 

\subsection{Stable mode when $Re_m\le100$} \label{sec:baseflow} 

\begin{figure}
    \centering
    \includegraphics[width=\linewidth]{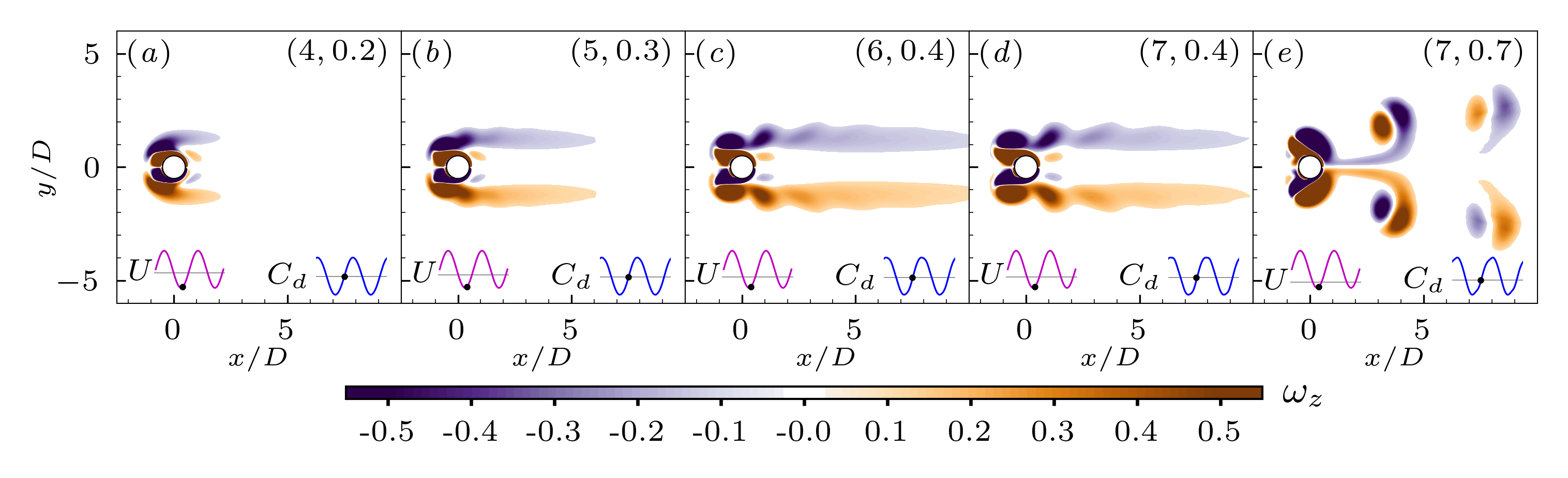}
    \caption{ Vorticity field of stable cases at $\Rey_m = 100$ and $\varphi = 1.60\pi$. Low-$m$ cases $(4,0.2)$, $(5,0.3)$, $(6,0.4)$  and $(7,0.4)$ show strong cycle-to-cycle vortex interaction with the cylinder; the higher-$m$ case $(7,0.7)$ show the shed structures carried downstream and away from the cylinder, with small cycle-to-cycle interaction. The corresponding free-stream velocity $U$ (\protect\magentaline) and drag coefficient $C_d$ (\protect\blueline) are inset.  
    }
    \label{fig:vorticity_stable}
\end{figure}

For the stable flow region (pale-blue background with 
$KC\lesssim6$) of figure~\ref{fig:stability_map}, the base flows identified by blue markers --in the absence of background hatching -- do not bifurcate when $Re_m=100$, nor do they at any lower value of $Re_m$.  
The associated re-stabilisation cases (pale-blue background with hatching), which are found to be unstable at a lower $Re_m$ are discussed in \S\ref{sec:island}. 
These stable flows are supported by the fully-developed nonlinear DNS results shown in figure~\ref{fig:vorticity_stable}, where the distribution of vorticity, $\omega_z=\nabla\times\mathbf{u}$, is displayed at a phase angle $\varphi=2\pi t/T=1.60\pi$, close to the instant of maximum flow reversal. Since every $\mathbf{u'}$ is damped, these are also examples of based flow amenable to Floquet analysis, satisfying $T$-periodicity and reflective symmetry, and consequently $C_l(t)=0$. 

During the forward half-cycle, $0\le \varphi\le \pi$, shear layers develop behind the cylinder, symmetric about $y=0$. As $U_m$ reverses during $\pi\le \varphi\le 2\pi$, the subsequent evolution of these shear layer is governed by the magnitude of $m$. According to (\ref{eqn:BCU}), fluid particles advance forward in the $+x$ direction over a distance:
\begin{eqnarray} 
    \frac{\Delta x^{+}}{D} = \frac{KC}{\pi} \left( m\pi+\sqrt{1-m^2}-m\arccos{m}\right),
    \label{eqn:+x}
\end{eqnarray}
and reverse in the $-x$ direction by:
\begin{eqnarray}
    \frac{\Delta x^{-}}{D} = \frac{KC}{\pi} \left( \sqrt{1-m^2}-m\arccos{m}\right).
    \label{eqn:-x}
\end{eqnarray}
For small $m$, such as those shown in figure~\ref{fig:vorticity_stable} (\textit{a}--\textit{b}), the reverse flow is strong enough to convect the shear layers developed under the effect of (\ref{eqn:+x}) 
backward into the $-x$ region, interacting with the newly formed shear layer induced by (\ref{eqn:-x}). As a result, cycle-to-cycle interaction is significant. For larger $m$, the influence of the reverse flow weakens rapidly; for example, $\Delta x^{-}/\Delta x^{+}\lesssim0.1$ for $m>0.6$. The interaction is therefore substantially attenuated. Consequently, the wake manifests itself as strongly discrete vortex pairs shed periodically in the $+x$ direction, as shown in figure~\ref{fig:vorticity_stable}(\textit{e}), rather than a pair of wavy vortex sheets as observed in figures~\ref{fig:vorticity_stable} (\textit{a}-\textit{d}).

\subsection{Pathways to a single unstable mode}\label{sec:routes}

\begin{figure}
    \centering
    \includegraphics[width=1\linewidth]{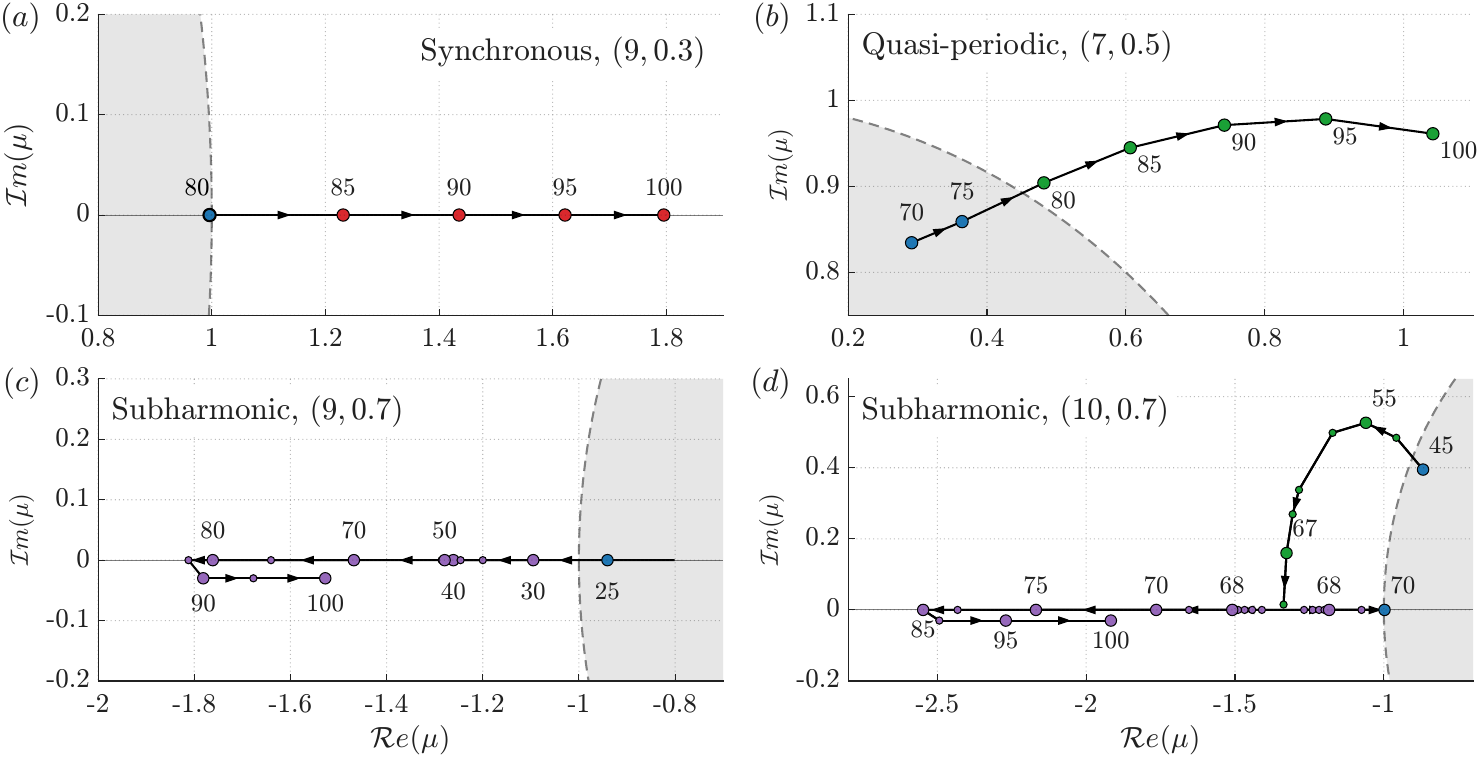}
    \caption{Loci of the leading Floquet multiplier $\mu_0$ for the given $(KC, m)$ condition as $\Rey_m$ increases from subcritical values to $\Rey_m=100$, with arrows indicating the direction of increasing $\Rey_m$.  Values of $\Rey_m$ are annotated alongside the loci, although not all values examined are labelled for the purposes of clarity. The dashed curve marks the boundary of the unit circle. Marker colouring follows the instability classifications in figure~\ref{fig:stability_map}. ($a$) the $S$ mode for $(KC, m)=(9, 0.3)$; $\mu_0$ exits the unit circle and stays as $\arg(\mu_0)=0$. ($b$) the $QP$ mode for $(7, 0.5)$: $\mu_0,\mu_1$ exit the unit circle and stay as a complex-conjugate pair (only $\mu_0$ is shown). ($c$) the $SH$ mode for $(9, 0.7)$: $\mu_0$ exits the unit circle and stays as $\arg(\mu_0)=\pi$. ($d$) for the case $(10, 0.7)$, the flow bifurcates as $QP$, with $\mu_0$ subsequently splitting on the $\arg(\mu_0)=\pi$ axis into two multipliers migrating in opposite directions, resulting in the $SH$ unstable mode at $Re_m=100$. Markers in the range $90\le\Rey_m\le100$ in (\textit{c}) and $85\le\Rey_m\le100$ in (\textit{d}) have been shifted slightly below the real axis for clarity. 
    }
    \label{fig:stability_pattern}
\end{figure}

To elucidate the 
nature of the unstable modes (the leading Floquet multiplier $|\mu_0|>1$) in figure~\ref{fig:stability_map}, the path/locus of $\mu_0$ in the complex plane is 
traced by stepping down $\Rey_m$, and hence $Re_p$, for a given $(KC, m)$ condition until the case becomes, and remains stable ($|\mu_0|<1$). 
Figure~\ref{fig:stability_pattern} is a collection of $\mu_0$ loci for four representative cases of single unstable modes present 
in figure~\ref{fig:stability_map}. Panels ($a$)--($d$) show the three primary types: the $S$ mode, the $QP$ mode, and two examples of the $SH$ mode, respectively.

\subsubsection{Synchronous ($S$) mode}\label{sec:syn}

\citet{Elston2006} found that a purely oscillatory flow ($m=0$) remains subcritical for $KC \leq 6$ and the onset of synchronous symmetry breaking to occur at $KC \ge 7$ (cf.\ figure~\ref{fig:elston_reproduce}). 
When $KC \approx 7.0$, the flow bifurcates -- becoming $QP$ before transitioning to being $S$,
in a manner delineated in figure~\ref{fig:elston_reproduce} (\textit{b}) as $\Rey_m$ increases. For $KC \ge 7.2$, the flow bifurcates to become $S$ directly.

When $U_c$ is superposed ($m>0$), all flows in the $S$ region of figure~\ref{fig:stability_map} bifurcate,
without becoming $QP$,
similar to that shown in figure~\ref{fig:stability_pattern}($a$), where the $\mu_0$ locus of the case $(9,0.3)$ is plotted as $\Rey_m$ increases from $80$ to $100$. 
It shows that the flow undergoes a canonical pitchfork bifurcation of the Poincar\'e map at $\Rey_m \approx 80$ \citep{iooss2012elementary} and the symmetry-breaking perturbation grows in phase with the external forcing, locking the emergent vortex shedding pattern to the fundamental forcing frequency $f_0 (=1/T)$. Thereafter, $\mu_0$ moves along the positive real axis until $\Rey_m$ reaches $100$ when $\mu_0\approx 1.796$.

\begin{figure}
    \centering
    \includegraphics[width=\linewidth]{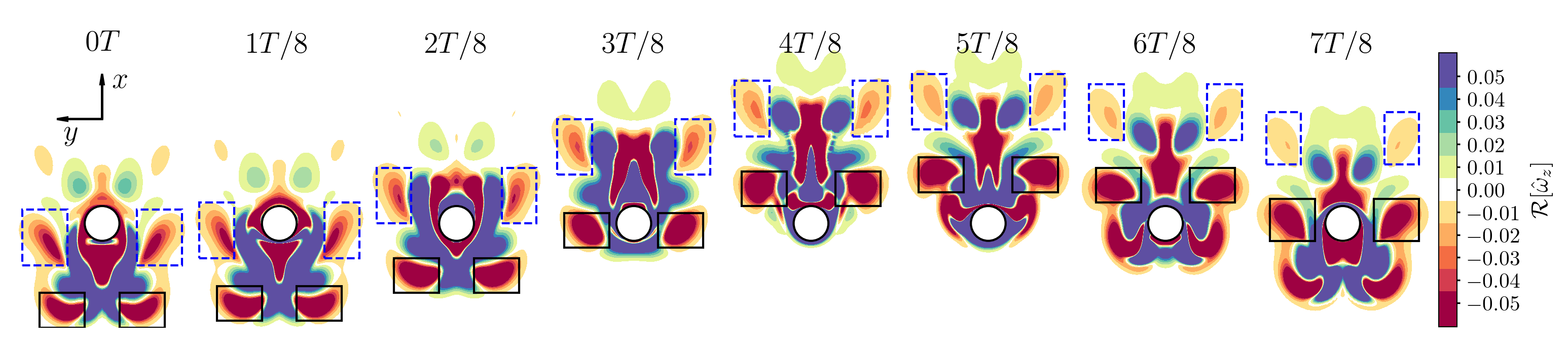}
    \caption{Phase-resolved evolution of the leading Floquet mode over one fundamental oscillation period $T$ for the case $(KC, m) = (9, 0.3)$ evaluated when $\Rey_m=100$. Contours show the real part of the vorticity mode, $\mathcal{R}e[\hat{\omega}_z]$. The solid and dashed lined boxes indicate the vortices generated from two consecutive cycles, respectively.
    }
    \label{fig:kc9m0p3M0}
\end{figure}

The leading Floquet mode evaluated at $\Rey_m = 100$ is presented in figure~\ref{fig:kc9m0p3M0}, which shows the real part of the vorticity mode $\mathcal{R}e[\hat{\omega}_z]$, where $\hat{\omega}_z=\nabla\times\hat{\mathbf{u}}$. The modal structure remaining from the previous cycle (highlighted by dashed boxes) after the action of the revered flow during the second half-cycle is trapped around the cylinder, while a pair of newly generated vortex packets (solid boxes) is formed. After a period, these structures advect with the bulk flow together in a very regular manner as the base flow oscillates, while their intensity increases. 

\begin{figure}
    \centering
    \includegraphics[width=1\linewidth]{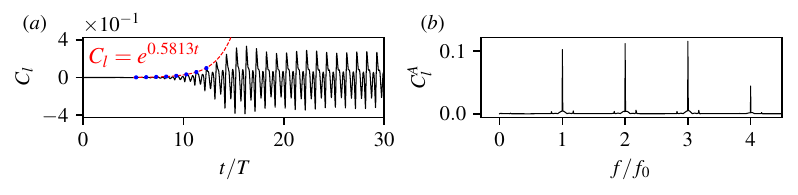}
	\caption{$C_l$ of the case $(KC, m) = (9, 0.3)$ evaluated when $\Rey_m=100$ by DNS. ($a$) Time history of $C_l$, fitted with an exponential function for its envelope under transient growth. ($b$) Frequency spectrum of $C_l$ computed after its oscillation saturates, showing dominant peaks at $f_0$ and its harmonics. 
    }
    \label{fig:ts_kc9_m0p3_S}
\end{figure}

DNS results at $\Rey_m = 100$ support the linear prediction, as demonstrated in figure~\ref{fig:ts_kc9_m0p3_S}($a$), which shows the time history of $C_l$ development. After an initial transient period, the envelope of $C_l$ grows exponentially before saturating into strictly $T$-periodic cycles. The exponential growth can be well approximated by $C_l=\exp(\sigma t)$, with the exponent coefficient $\sigma \approx 0.581$, in agreement with the value of the Floquet exponent, i.e., $\exp(0.581 T)=|\mu_0| \approx 1.788$.
The spectrum of $C_l$ computed over $200T \le t \le 300T$, with $C_l^A$ representing the corresponding power spectral density, is shown in figure~\ref{fig:ts_kc9_m0p3_S} ($b$), where dominant peaks are located at the fundamental frequency $f_0$ and its harmonics.  

\begin{figure}
    \centering
    \includegraphics[width=1\linewidth]{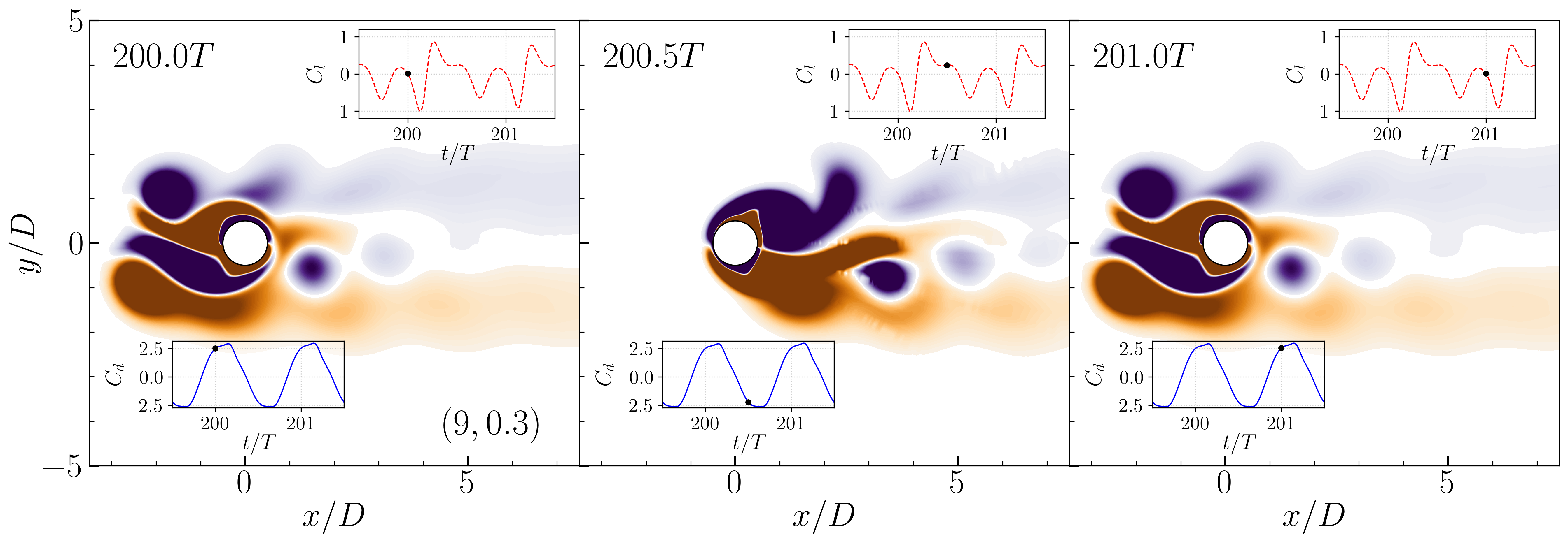}
    \caption{Fully developed $\omega_z$ distribution for the case $(KC, m) = (9, 0.3)$ evaluated when $\Rey_m=100$ by DNS, at $\varphi=0,\pi$ and $2\pi$. Colour-map key as in figure~\ref{fig:vorticity_stable}. The corresponding lift coefficient $C_{l}$ (\protect\magentaline) and drag coefficient $C_d$ (\protect\blueline) are inset.}
    \label{fig:kc9m0p3_dns_vorticity}
\end{figure}

The fully developed vortex pattern shown in figure~\ref{fig:kc9m0p3_dns_vorticity} confirms that this condition saturates the flow to $T$-periodicity but with a broken reflective symmetry. At $t=200T$ (phase $\varphi=0$), vortex structure is located mainly in the $-x$ region and is deflected to the $+y$ side. This flow pattern repeats exactly after $T$. In between, at $t=200.5T$ ($\varphi=\pi$), vortex structure moves to the $+x$ region, but is also deflected towards the same $+y$ direction. This net deflection of the vortex structure within a cycle is associated with a negative time-averaged lift $\overline{C_l}$, as confirmed by the inset $C_l$ panels. 
The direction of wake deflection depends on the initial condition, which is stochastic. Flipping the flow field to generate a new initial condition yields a mirrored solution, consistent with structures of a supercritical pitchfork bifurcation.

\subsubsection{Quasi-Periodic (QP) mode}\label{sec:qp}

Figure~\ref{fig:stability_pattern}($b$) shows the locus of the leading multipliers as $\Rey_m$ increases. For this particular case $(KC, m) = (7, 0.5)$, the flow bifurcates, between $\Rey_m = 75$ and $80$, via a Neimark--Sacker (secondary Hopf) type \citep{iooss2012elementary}, featuring a complex-conjugate pair of multipliers that cross the unit circle. The $T$-periodic base flow loses stability and develops into a $QP$ flow on a $T^2$ invariant torus, characterised by a secondary frequency denoted as $f_s^\mathcal{L}$ (with $\mathcal{L}$ standing for linear framework), which can be determined by the argument of the critical multiplier, $f_s^\mathcal{L}/f_0 = \arg(\mu_0)/2\pi \approx 0.1721$ at onset, decreasing to $0.119$ at $\Rey_m = 100$.

\begin{figure}
    \centering
    \includegraphics[width=1\linewidth]{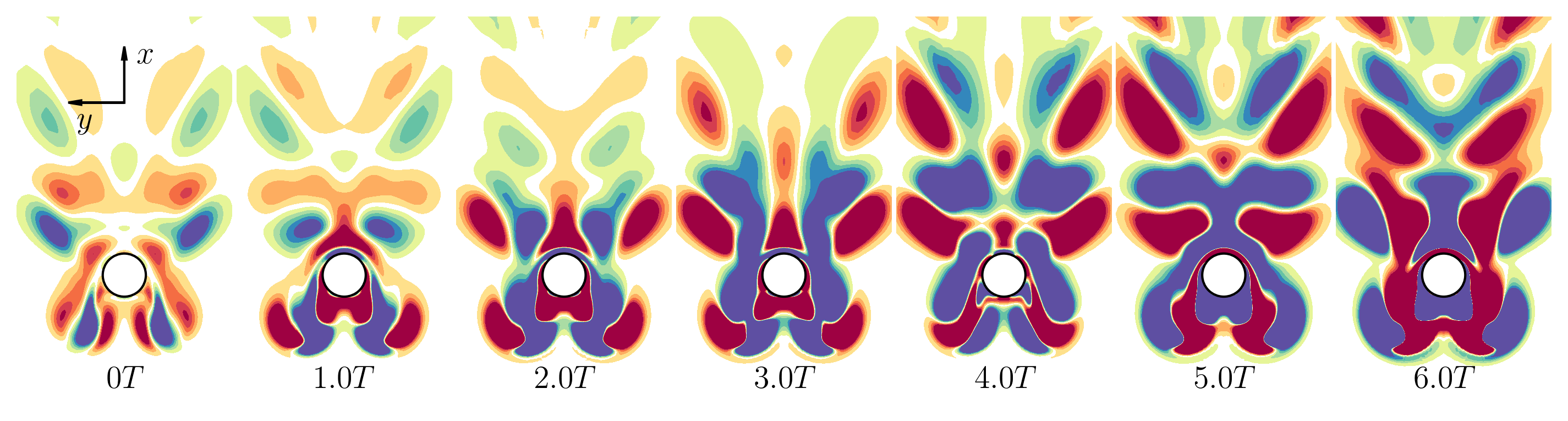}
    \includegraphics[width=1\linewidth]{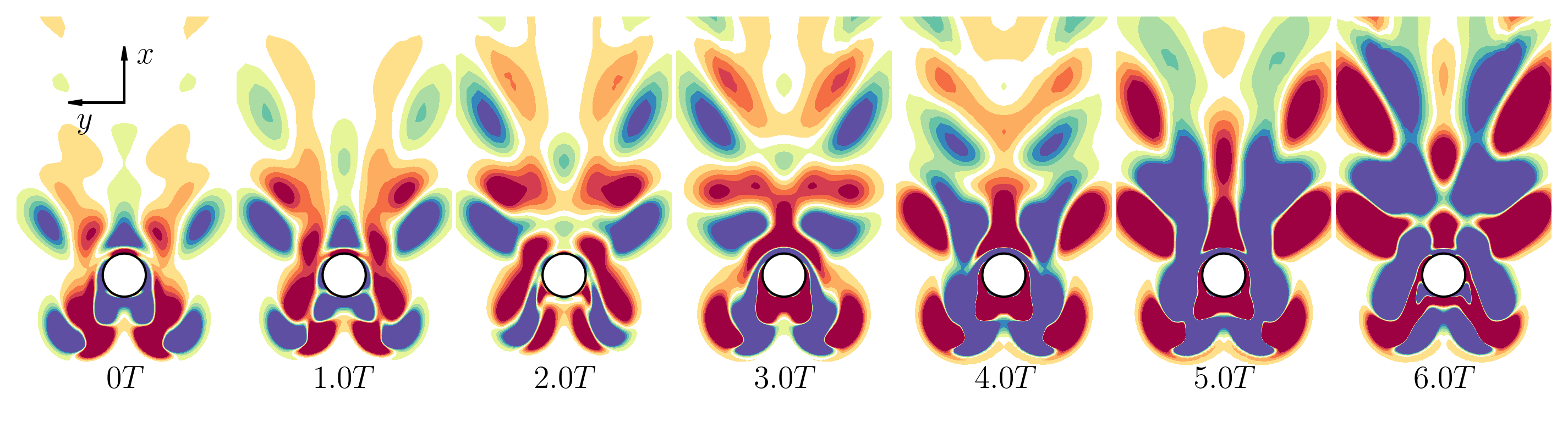}
    \caption{Evolution of $\mathcal{R}e[\hat{\omega}_z]$ (top row) and $\mathcal{I}m[\hat{\omega}_z]$ (bottom row) of the leading Floquet mode for $(KC, m) = (7, 0.5)$ when $\Rey_m = 100$, shown over six forcing periods. The two rows share the same spatial structure but are shifted in phase for $2T$. Colour-map key as in figure~\ref{fig:kc9m0p3M0}.}
    \label{fig:kc7m0p5QP_real}
\end{figure}

Figures~\ref{fig:kc7m0p5QP_real} shows the real and imaginary parts of the corresponding Floquet mode $\hat{\omega}_z$, evolved over $6T$. With the intensity clearly increasing over successive cycles, the two components share a similar spatial structure but are phase-shifted by $2T$: $\mathcal{R}e[\hat{\omega}_z]$ at $t = 1.0T$, $2.0T$, $3.0T$... closely resembles $\mathcal{I}m[\hat{\omega}_z]$ at $t = 3.0T$, $4.0T$, $5.0T$... This corroborates the rotation of a complex-conjugate pair on the torus, with the amplitude growing over the post-bifurcation interval.

\begin{figure}
    \centering
    \includegraphics[width=1\linewidth]{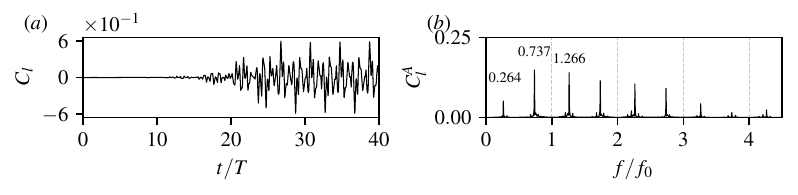}
    \caption{$C_l$ of the case $(KC, m) = (7, 0.5)$ evaluated when $\Rey_m=100$ by DNS.  (\textit{a}) $C_l$ development in time-domain. (\textit{b}) Frequency spectrum of $C_l$ computed after its oscillation saturates, showing peaks at $n \pm \Delta f/f_0$, where the incommensurate secondary frequency $f_{s}^{\mathrm{DNS}}=\Delta f \approx 0.264f_0$.}
    \label{fig:ts_kc7_m0p5_QP}
\end{figure}

DNS simulation at $\Rey_m = 100$ confirms flow quasi-periodicity, supported by the time domain $C_l(t)$ for $t>30T$, figure~\ref{fig:ts_kc7_m0p5_QP} (\textit{a}), and the spectrum of $C_l(t)$ computed over $t/T_0 \in [200,300]$ which shows peaks located at $n \pm \Delta f/f_0$ with $n \in \mathbb{N}$, figure~\ref{fig:ts_kc7_m0p5_QP} (\textit{b}). The secondary frequency extracted from this nonlinearly saturated flow can be determined as $f_{s}^{\mathrm{DNS}}=\Delta f = 0.264f_0$, which is also the location of the first peak of the spectrum. It is apparent that $f_{s}^{\mathrm{DNS}}$ is markedly higher than the linear prediction $f_s^\mathcal{L} = 0.119f_0$ at the same $\Rey_m$. 
A comparable upward shift of the secondary frequency, from its linear estimate, has been reported for bifurcated purely oscillatory flow \citep{Elston2006}. This is due to the nonlinear effect of a kind similar to  bifurcated purely oscillatory flow and the result of accumulated phase mismatch after many cycles.
The $QP$ attractor plays a central role in the nonlinear dynamics of the coexistence regime treated in \S\ref{sec:coexist}, where it persists in the saturated wake even under conditions in which it is not linearly dominant.

\subsubsection{Subharmonic (SH) mode}\label{sec:sub}

The new subharmonic mode, forbidden for a $Z_2$-symmetric limit cycle according to \cite{Swift1984}, and indeed absent in \cite{Elston2006} over the same $KC$ range at $m=0$, can be unlocked when the steady component $U_c$ becomes sufficiently large, i.e. for $m\gtrsim0.6$. 
The two representative cases at $m=0.7$ illustrated in figure~\ref{fig:stability_pattern}($c$) and ($d$) demonstrate how the $SH$ channel can be accessed. For the former, $(KC, m) = (9, 0.7)$, the flow bifurcates at $Re_m\approx27$ ($Re_p\approx46$) as the leading multiplier exits the unit circle at $\mu_0 = -1$. Subsequently, $\mu_0$ stays on the negative real axis, i.e. $\arg(\mu_0)=\pi$ and a period-doubling mode persists until $\Rey_m=100$, without other modes emerging at any $\Rey_m$ between. The Floquet mode exactly inverts sign after one forcing period $T$ \citep{iooss2012elementary}, and the shedding frequency is locked at exactly $f_s/f_0 = 0.5$ from the DNS result,
independent of $\Rey_m$ within the range. Somewhere in $80<\Rey_m<90$, the $\mu_0$ locus makes a U-turn where $\mu_0\approx-1.8$, and subsequently it migrates towards the unit circle. At $\Rey_m=100$, $\mu_0=-1.53$, 
which suggests, interestingly, that the flow destabilises more gradually than within the range $73\lesssim\Rey_m<100$.

The case $(KC, m) = (10, 0.7)$ shown in figure~\ref{fig:stability_pattern}($d$) displays a qualitatively different locus of the leading multiplier. The flow bifurcates at $\Rey_m \approx 47.5$ ($\Rey_p \approx 81$) as $QP$ first. The unstable mode then transitions to $SH$ 
at $\Rey_m \approx 67.6$ ($\Rey_p \approx 115$), when the complex-conjugate pair coalesces on the negative real axis at $\mu_0\approx-1.337$ 
before splitting into two real negative multipliers ($\mu_0$ and $\mu_1$), similar to the formation of the $S$ mode shown in figure~\ref{fig:elston_reproduce} (\textit{b}). As $\Rey_m$ increases further, $|\mu_1|$ reduces, and re-enters the unit circle at $\Rey_m=70$, whereas $|\mu_0|$ continues to increase, moving further along the negative real axis and rendering the flow more unstable. It reaches a maximum at $\Rey_m\lesssim85$, after which the locus makes a U-turn and its magnitude reduces again, reaching $\mu_0 = -1.92$ at $\Rey_m = 100$. Another noticeable observation is that bifurcation occurs at a much higher $\Rey_m$ when $(KC, m) = (10, 0.7)$ than for $(9, 0.7)$ discussed above, despite the larger $KC$ at the same $m$, which would otherwise suggest a more `energetic' flow condition.

\begin{figure}
    \centering
    \includegraphics[width=1\linewidth]{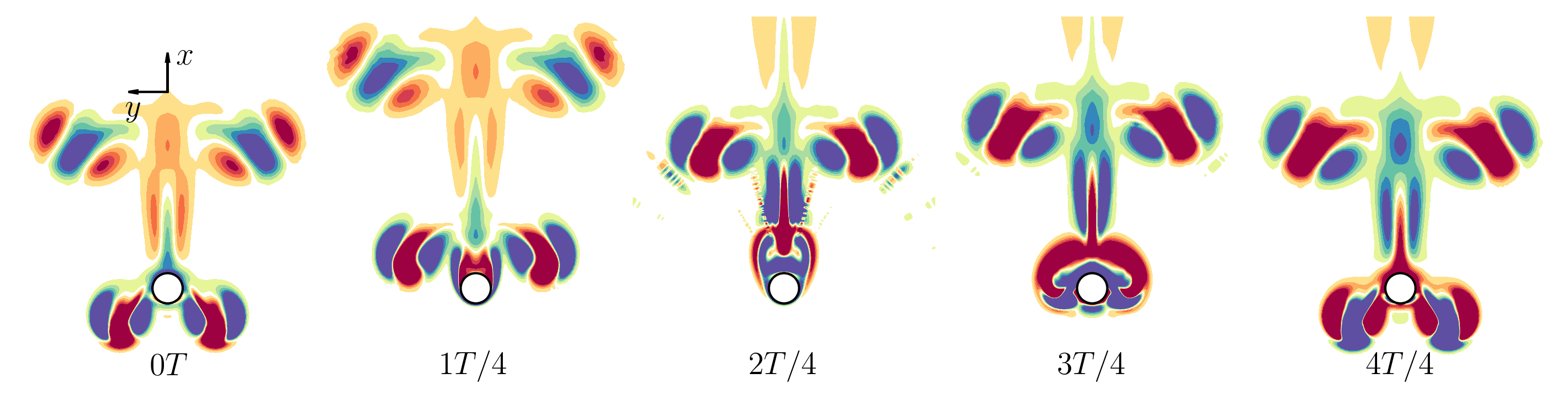}
    \caption{Phase-evolution of $\mathcal{R}e[\hat{\omega}_z]$ for case $(KC, m) = (10, 0.7)$ evaluated at $\Rey_m=100$ over one oscillation cycle. The identical structure, differing only by a sign inversion and intensity between $t$ and $t+T$, is a characteristic feature of period-doubling. Colour-map key as in figure~\ref{fig:kc9m0p3M0}.}
    \label{fig:kc10m0p7SH}
\end{figure}

The leading Floquet modes and fully developed DNS results for $(10, 0.7)$ when $\Rey_m = 100$ are now examined; those for $(9, 0.7)$ are qualitatively similar at this $\Rey_m$ and therefore are not presented separately. The phase-evolution of the Floquet mode for $(10, 0.7)$ illustrated in figure~\ref{fig:kc10m0p7SH} exhibits a period-doubling signature: sign-inverted modal distribution after one forcing period, viz., $\hat{\omega}_z(t + T) = -\hat{\omega}_z(t)$, with intensity continuing to grow, so that fully repeated (structure-wise) modal distribution requires $2T$ to achieve. Due to a larger $m$, compared to figure~\ref{fig:kc9m0p3M0} and \ref{fig:kc7m0p5QP_real}, modal structures generated from successive cycles are further apart downstream ($x>0$), and therefore their interaction is significantly reduced.

\begin{figure}
    \centering
    \includegraphics[width=1\linewidth]{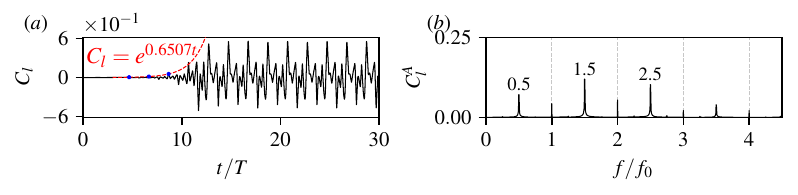}
    \caption{$C_l$ of the case $(KC, m) = (10, 0.7)$ evaluated when $\Rey_m=100$ by DNS. (\textit{a}) $C_l$ development in time-domain, showing alternating speaks repeating in $2T$. (\textit{b}) Frequency spectrum of $C_l$ computed after its oscillation saturates, with the dominant peak at the period-doubled frequency $f/f_0 = 0.5$.
    }
    \label{fig:ts_kc10_m0p7_sub}
\end{figure}

\begin{figure}
    \centering
    \includegraphics[width=1\linewidth]{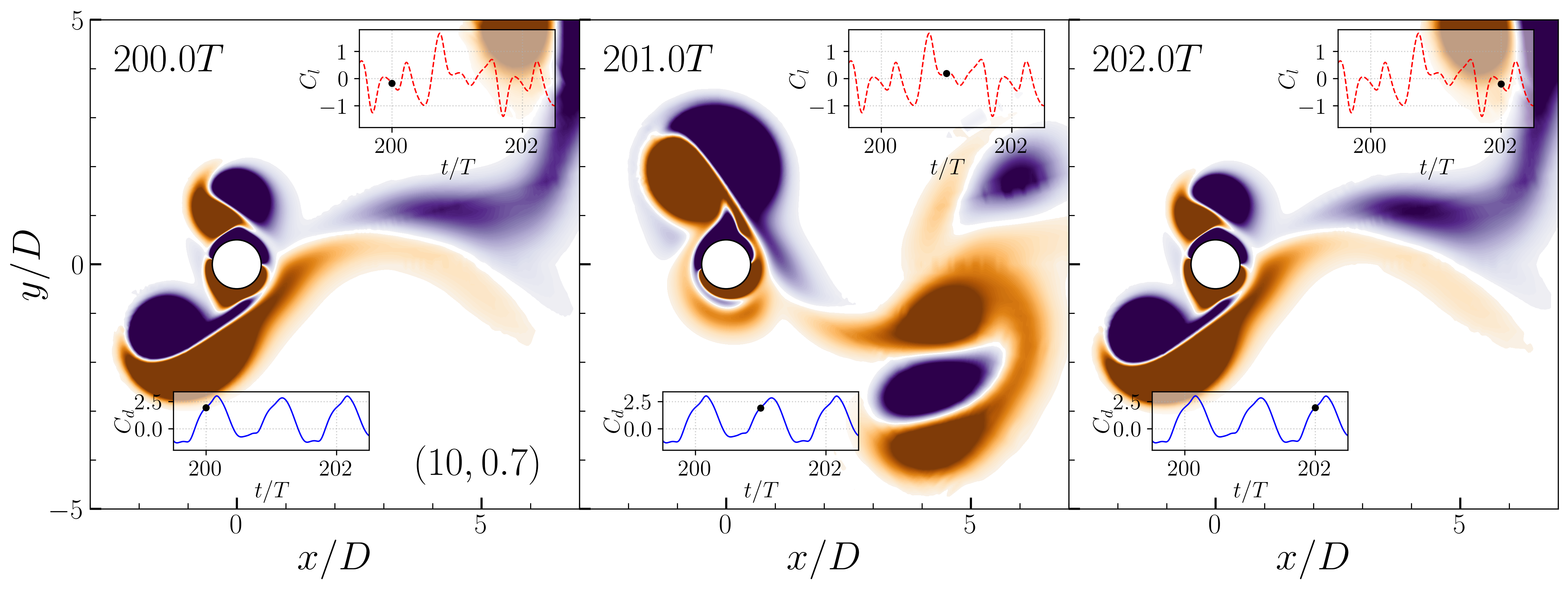}
    \caption{Fully developed $\omega_z$ contours for $(KC, m) = (10, 0.7)$ evaluated at $\Rey_m=100$ by DNS, showing perfect period-$2$ phase-locking over $2T$ intervals. Colour-map key as is figure~\ref{fig:vorticity_stable}. The corresponding lift coefficient $C_{l}$ (\protect\magentaline) and drag coefficient $C_d$ (\protect\blueline) are inset.}
    \label{fig:ts_kc10_m0p7_sub_vorticity}
\end{figure}

Results for $C_l$ from DNS simulation confirm the period-doubled state, figure~\ref{fig:ts_kc10_m0p7_sub}. The fully developed $C_l$ spectrum is clearly dominated by a primary peak at half the forcing frequency ($f=0.5f_0$), accompanied by odd half-integer harmonics ($1.5f_0, 2.5f_0, \ldots$). An exponential fit to the transient growth period of $C_l(t)$ gives $\sigma \approx 0.651$, corresponding to $\exp(\sigma T) \approx 1.92$, in agreement with the prediction from the linear analysis. That is, $\mu_0 = -1.92$ at $\Rey_m = 100$. The nonlinearly saturated wake shown in figure~\ref{fig:ts_kc10_m0p7_sub_vorticity} also confirms the $2T$-period structure in physical space: snapshots after $2T$ recover themselves identically. 

\subsection{Re-stabilised mode}\label{sec:island}
Figure~\ref{fig:stability_map} also reveals a second neutral stability curve for $KC\gtrsim7$ and $m\gtrsim0.9$ ($\Rey_p\approx190$), separating a second pale-blue stable region, from the adjacent pale-purple SH one, as indicated by the hatched lines at the top right of the map. 
As noted in \S\ref{sec:globalmap}, $\Rey=190$ exceeds the primary critical $\Rey$ for purely oscillatory flow and even the secondary critical $\Rey$ for a purely steady current past a circular cylinder. This striking restoration of the reflective symmetry therefore demonstrates once again that the $m>0$ condition departs markedly from a simple linear superposition of the two individual components. 
Additionally, there is a lower-left sub-region within the pale-blue stable range $KC\lesssim6$, hatched and adjacent to the pale-purple SH one, where flow has been through an unstable state at lower $\Rey_m$. The formation mechanisms for both of these two re-stabilised regions are analysed in turn below.

\begin{figure}
    \centering
    \includegraphics[width=0.9\linewidth]{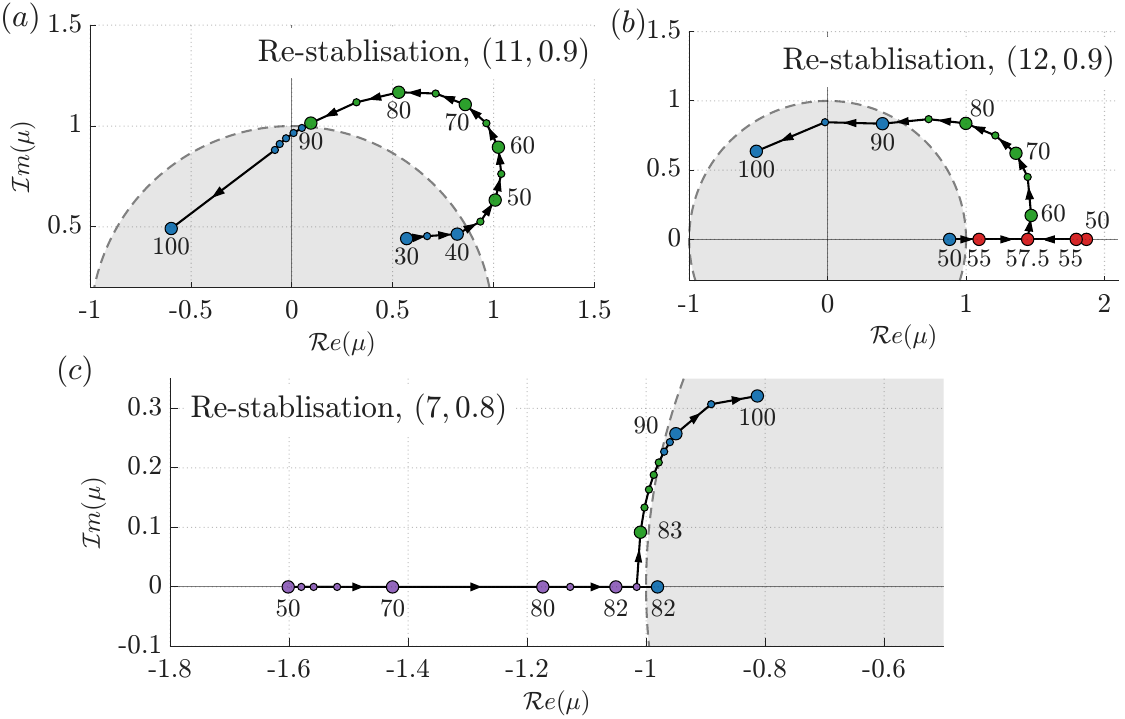}
    \caption{Loci of the leading Floquet multiplier $\mu_0$ for representative re-stabilisation cases. ($a$): pathway $QP\to\mathrm{re\text{-}stabilisation}$, $(KC,m)=(11, 0.9)$; ($b$): $S\to QP\to\mathrm{re\text{-}stabilisation}$, $(12, 0.9)$; ($c$): $SH\to QP\to\mathrm{re\text{-}stabilisation}$, $(12, 0.9)$. Arrows indicating the direction of increasing $\Rey_m$ to $100$. Marker colouring follows the instability classifications in figure~\ref{fig:stability_map} and the dashed curve marks the boundary of the unit circle. 
    }
    \label{fig:restable_pattern}
\end{figure}

Figure~\ref{fig:restable_pattern} 
illustrates the prevailing pathways leading to flow re-stabilisation in the upper-right region of figure~\ref{fig:stability_map}. Comparing the first two pathways which are both at $m=0.9$, case (\textit{a}), $KC=11$, presents a comparatively simpler $\mu_0$ locus. The base flow Neimark--Sacker bifurcation at $\Rey_m \approx 43$ becomes linearly unstable and 
$QP$. As $\Rey_m$ is increased further $|\mu_0|$ grows, reaching a maximum at $\Rey_m \approx 70$ and beyond which it diminishes, eventually re-entering the unit circle at $\Rey_m \approx 90$. Increasing $\Rey_m$ at this fixed $(KC, m)$ therefore stabilises a configuration that is linearly unstable at lower $\Rey_m$. The flow then remains stable for $90 < \Rey_m \leq 100$.

On increasing $KC$ by 1, the locus of $\mu_0$ for ($12,0.9$) differs markedly, figure~\ref{fig:restable_pattern} (\textit{b}), which is the only exceptional case in the upper-right sub-region. At $\Rey_m = 50$ ($\Rey_p = 95$), the flow is already synchronous and unstable, with $\mu_0=1.87$.  
As $\Rey_m$ is increased, the leading multiplier $\mu_0$ reduces, suggesting the base flow is less susceptible to instability. Concurrently, a second multiplier of the same type, $\mu_1$, exits the unit circle at $+1$ and becomes sub-leading. Both $\mu_0$ and $\mu_1$ then migrate towards each other along the real axis until they coalesce when $\Rey_m \approx 57.5$. This 
$S+S$ coalescence, 
which will be elaborated on in \S\ref{sec:coexist}, 
transitions to become $QP$.
At $\Rey_m \approx 87$ the pair re-enters the unit circle, and flow re-stabilises, remaining stable at $\Rey_m = 100$. 

Figure~\ref{fig:restable_pattern} (\textit{c}) depicts an example pathway to re-stabilisation for the cases in the lower-left pale-blue hatched region of figure~\ref{fig:stability_map}, which adjoins the `always' stable cases discussed in \S\ref{sec:syn}. From an intermediate $\Rey_m=50$ up to $100$, the $\mu_0$ locus follows a trajectory similar to that shown in figure~\ref{fig:stability_pattern} (\textit{d}), but in the reverse direction and via a more subtle pathway in general. In particular, the sequence $SH\to QP\to\mathrm{re\text{-}stabilisation}$ spans the narrow range $82<\Rey_m<88$, with $\mu_0$ in the $QP$ range moving almost tangentially along the unit circle from $\Rey_m=85$ to $88$ where it finally enters the circle at $\mu_{0,1}=-0.978\pm 0.209i$.

\begin{figure}
    \centering
    \includegraphics[width=1\linewidth]{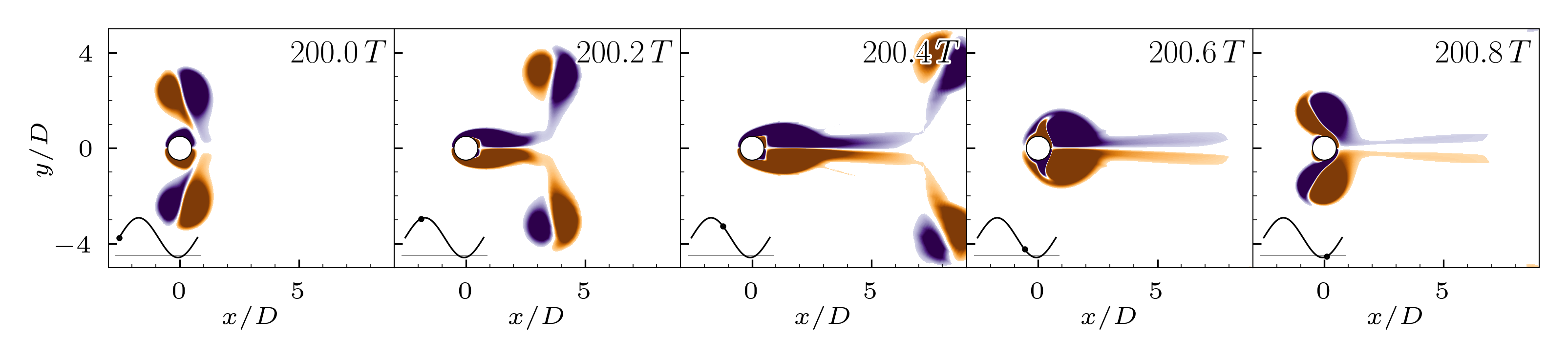}
    \caption{Cycle variation of fully developed $\omega_z$ contour for $(KC, m) = (12, 0.9)$, $\Rey_m=100$. The flow field remains symmetric with respect to the wake centreline. Colour-map key as in figure~\ref{fig:vorticity_stable}.}
    \label{fig:vorticity_kc12m0p9_stable}
\end{figure}

The mechanism underlying such flow re-stabilisation is partially illuminated by observations in physical space using the DNS results presented in figure~\ref{fig:vorticity_kc12m0p9_stable}, together with a comparison of the forward and reverse convection length scales within a single forcing period. According to equations~(\ref{eqn:+x}) and (\ref{eqn:-x}), as the ratio $\Delta x^{-}/\Delta x^{+}$ diminishes to about $0.02$ for $m =0.9$, the action of the reversed phase is primarily to promote the roll-up of coherent vortex-packet pairs from the shear layers, and expel the vortices formed in the previous cycle transversely, with only a negligible displacement in the $-x$ direction. Inter-cycle vortex interaction is therefore greatly reduced, which facilitates the formation of distinct vortex pairs and well-organised shear layers that are convected efficiently and symmetrically about the wake centreline in the $+x$ direction. 

At $\Rey_p=190$, the initial growth rate of a small perturbation in a steady current past a circular cylinder can be estimated from a global modal stability framework as $\sigma_\mathcal{H}D/U\sim\mathcal{O}(10^{-1})$ 
\citep{Barkley2006}, 
 while recognising that nonlinear effects may already be significant since this Reynolds number lies well above the critical threshold for the Hopf bifurcation ($\Rey\approx46$). According to (\ref{eqn:+x}), the flow condition ($KC\geq11,m=0.9$) yields a unidirectional convection distance of $\Delta x^+/D\sim\mathcal{O}(10)$ within a cycle, which is comparable to that associated with $\sigma_\mathcal{H}$. Therefore, it seems that a comparatively very small reverse direction convection $\Delta x^-/D\sim\mathcal{O}(10^{-1})$, and/or the unsteady nature of the $U_m$ component, together with the coupling effect of $U_c$ and $U_m$ of course, acts to suppress the development of a K\'arm\'an type of spatio-temporal instability within a cycle.

\subsection{Mode coexistence}\label{sec:coexist}

In figure~\ref{fig:stability_map}, there is also a pale-yellow shaded region in which the markers indicate mode combinations of $S$+$QP$ or $S$+$S$. Occupying the parameter range $KC \geq 10$ and $0.1 < m \leq 0.4$, this region is characterised by the coexistence of two instability modes when $\Rey_m=100$, corresponding to two Floquet multipliers lying outside the unit circle.
$S$+$S$ coexistence  has also been shown to occur at lower $\Rey_m$ for all $S+QP$ cases; see figure~\ref{fig:restable_pattern}(\textit{b}). 

\begin{figure}
    \centering
    \includegraphics[width=1\linewidth]{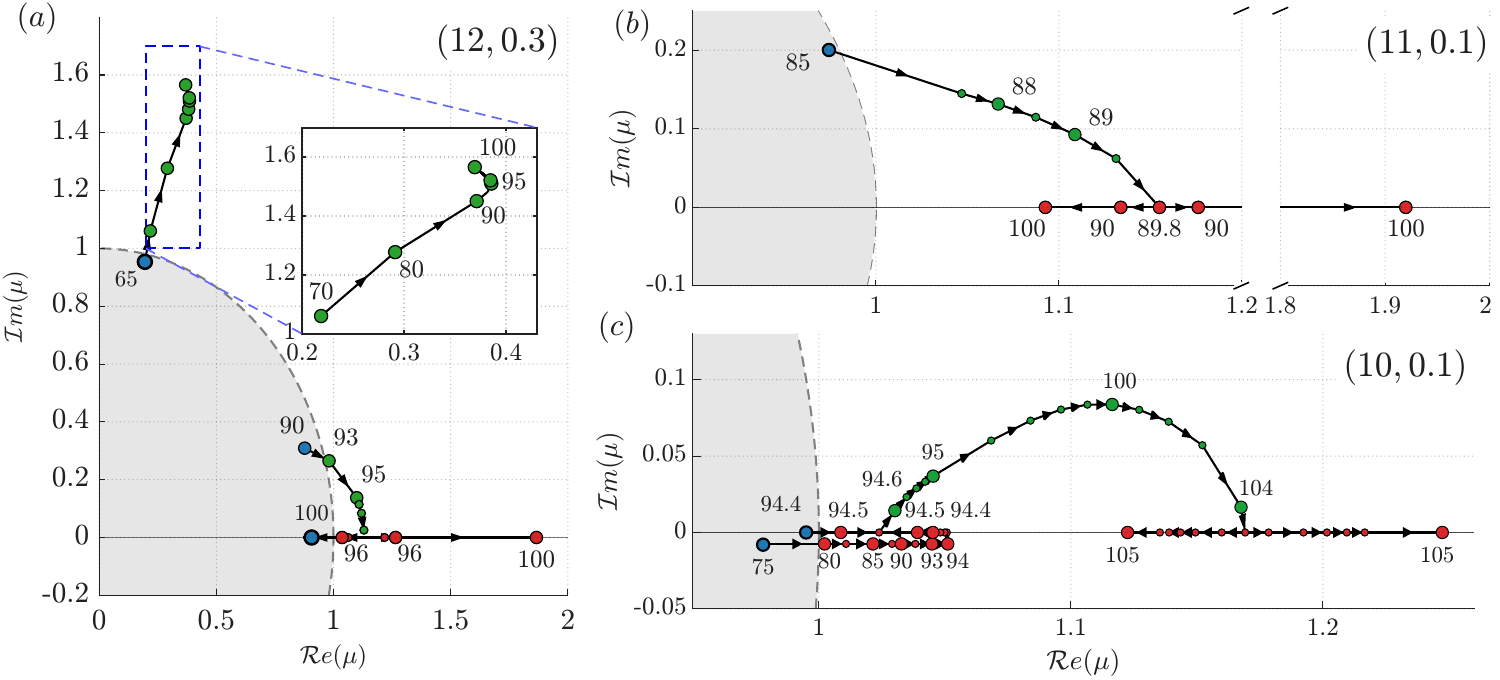}
    \caption{Loci of the leading Floquet multipliers in the complex plane for three cases as $\Rey_m$ increases at given $(KC, m)$ condition. Labels give $\Rey_m$ values; arrows show the direction of increasing $\Rey_m$.  ($a$) $(12, 0.3)$: mode coexisting $S+QP$ (the complex-conjugate locus is not shown). ($b$) $(11, 0.1)$: mode coexisting $S+S$. ($c$) $(10, 0.1)$: a special transitional case $QP$ at $\Rey_m = 100$. The range $75\le\Rey_m\le94$ is shifted slightly below the real axis for clarity. Marker colouring follows the instability classifications in figure~\ref{fig:stability_map} and the dashed curve marks the boundary of the unit circle.}
    \label{fig:floquet_compiled_coexist}
\end{figure}

In principle, both unstable modes can trigger the exponential growth of infinitesimal perturbations individually, although linear stability theory predicts that the mode associated with the larger Floquet multiplier magnitude will dominate the perturbation asymptotically during the linear growth phase \citep{Schmid2001}. On closer examination the fully developed DNS results for these conditions reveal, however, that there is evidently a non-straightforward competition between the two unstable modes, plausibly via nonlinear interactions. This is demonstrated by the case $(KC, m) = (12, 0.3)$ shown in figure~\ref{fig:floquet_compiled_coexist}(\textit{a}), where two branches for different unstable modes interact.

\begin{figure}
    \centering
    \includegraphics[width=1\linewidth]{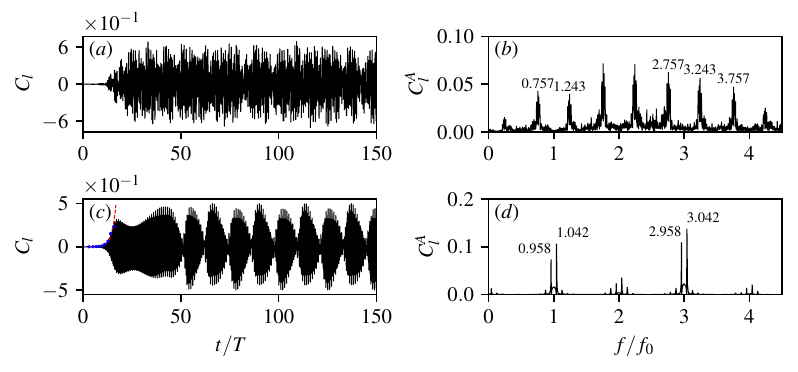}
    \caption{ DNS time history of the lift coefficient $C_l$ and its frequency spectrum at $\Rey_m = 100$. (\textit{a},\textit{b}) $(KC, m) = (12, 0.3)$, with quasi-periodic peaks at $f/f_0 = n \pm 0.243$; (\textit{c},\textit{d}) $(KC, m) = (11, 0.1)$, with peaks at $f/f_0 = n \pm 0.042$, $n \in \mathbb{N}$. 
    }
    \label{fig:ts_kc12_m0p3_kc11_m0p1_co}
\end{figure}

This flow  first becomes unstable through a Neimark--Sacker bifurcation 
becoming
$QP$ at $Re_m \approx 67$, where a complex-conjugate pair of multipliers crosses the unit circle. The magnitude of this pair grows with $\Rey_m$ while the flow retains its $QP$ character. Strikingly, as $\Rey_m$ approaches $93$, a second instability sets in: another complex-conjugate pair with smaller argument leaves the unit circle. This pair migrates towards the positive real axis, where the two multipliers coalesce and split into a pair of real multipliers moving in opposite directions over the range $95<\Rey_m<96$. The resulting branches resemble the scenario illustrated in figure~\ref{fig:elston_reproduce}(\textit{b}), except that here the two unstable modes coexist. At $\Rey_m=100$, the flow therefore manifests two unstable modes: a leading 
$S$ one with $\mu_0=1.866$, and a sub-leading 
$QP$ one with $\mu_{1,2}=0.3693\pm 1.566i$ ($|\mu_{1,2}|\approx1.61$), together with a second $S$ mode being stable $\mu_3=0.9066$. 
Since $|\mu_0|>|\mu_{1,2}|$, linear theory predicts that $S$ grows more rapidly and would dominate the initial linear development stage of the perturbation. This contrasts with the DNS results in figure~\ref{fig:ts_kc12_m0p3_kc11_m0p1_co} (\textit{a}--\textit{b}), which display QP behaviour that persists into the nonlinear saturated stage. Strongly irregular amplitude modulation is noticed in (\textit{a}), indicating the interaction between two modes. This interpretation is supported by the QP structure of the spectrum in (\textit{b}), in which spectral peaks are observed at $f/f_0 = n \pm 0.243$, with $n \in \mathbb{N}$; see \S\ref{sec:qp}. 
The spectrum also features a broader band of low intensity spanning a wide range of frequencies, suggesting an overall more chaotic flow field (figure not shown), plausibly attributable to nonlinear mode interaction.

The formation mechanism of the other mode coexistence type, $S$+$S$, is illustrated for the case $(KC, m) = (11, 0.1)$ in figure~\ref{fig:floquet_compiled_coexist} (\textit{b}). Similar to the case in (\textit{a}), the base flow bifurcates becoming $QP$ at $\Rey_m\approx86$, although the argument of the associated complex-conjugate pair is appreciably smaller than that in (\textit{a}).
The pair then follows a familiar trajectory towards the real axis, coalescing at $\Rey_m=89.8$ before splitting into two real and positive multipliers that subsequently migrate towards the $+x$ and $-x$ directions, respectively. At $\Rey_m=100$, $\mu_0=1.9198$ and $\mu_1=1.0926$ 
indicating the coexistence of two $S$ modes both `locked' onto the fundamental forcing frequency $f_0$, unlike the case shown in figure~\ref{fig:floquet_compiled_coexist} (\textit{a}), where a secondary unstable 
$QP$ branch emerges up to $\Rey_m=100$.

The time history of $C_l(t)$ shown in figure~\ref{fig:ts_kc12_m0p3_kc11_m0p1_co}(\textit{c}), obtained from DNS 
for $\Rey_m=100$, clearly demonstrates that the $S$+$S$ unstable modes dictate the transient growth at early time $t<25T$, where $C_l$ varies synchronously with the fundamental frequency $f_0$ and an exponential fit agrees with the growth rate at small time. 
Nevertheless and remarkably, despite the absence of an unstable $QP$ mode, the long term fully developed flow field features clear amplitude modulation starting at $t\approx60T$. This behaviour is further supported by the spectrum shown in (\textit{d}), evaluated over the fully developed interval $t > 100T$,  
where spectral peaks are identified at $f/f_0 = n \pm 0.042$. 
This breaking of synchronisation -- albeit weak -- could, in addition to the subtle nonlinear interaction between the two coexisting $S$ modes of different multiplier magnitudes, also be associated with the lower instantaneous Reynolds number attained within each cycle according to equation (\ref{eqn:BCU}) permitting a $QP$ mode to manifest and gradually accumulate over many cycles. It is unlikely, however, to arise from $\Rey_m$ hysteresis, since the flow evolution does not pass through a lower $\Rey_m$ state which is $QP$ unstable.
The subtle nature of this synchronous symmetry breaking is further reflected in the other $S{+}S$ case, $(12,0.1)$: with only a marginal increase of $KC$ by 1, the spectrum of $C_l$ indicates purely synchronous symmetry breaking (figure not shown), albeit with $\mu_0,\mu_1$ pathways similar to that shown in figure~\ref{fig:floquet_compiled_coexist} (\textit{a}) (the lower branch).

The fully developed DNS simulations for the two cases at $\Rey_m=100$ therefore show that the coexisting unstable modes identified within the linear Floquet framework, while governing the initial transient growth, unlike the single mode scenarios discussed in \S\ref{sec:routes}, do not necessarily shape the nonlinear saturated state, where the flow appears to favour the quasi-periodic dynamics. This is observed for all mode coexistence cases at $\Rey_m=100$, except $(12,0.1)$.

Keeping the same $m$ while reducing $KC$ by 1 leads to the exceptional case $(KC,m)=(10,0.1)$, corresponding to the isolated green marker surrounded by red markers in figure~\ref{fig:stability_map}. Also located within the mode coexistence region, the formation of the unstable $QP$ mode is elucidated here. Unlike the cases discussed in \S\ref{sec:qp}, although this case shows single $QP$ mode instability 
at $\Rey_m=100$, the locus of its leading Floquet multipliers traces a unique and rather intricate pathway, as can be seen in figure~\ref{fig:floquet_compiled_coexist} (\textit{c}). The $Z_2$ symmetry is broken by the emergence of a single synchronous unstable mode with $\mu_0=1$ at $Re_m\approx80$. The real and positive $\mu_0$ subsequently increases with $\Rey_m$ until the sub-leading real multiplier $\mu_1$ exits the unit circle at $94.4<\Rey<94.5$, marking the onset of the coexistence of two unstable synchronous modes $S$+$S$. At this instant, $\mu_0$ starts to migrate backwards and $\mu_1$ forwards, and very quickly, they merge and split into a pair of complex-conjugate multipliers as $\Rey_m$ increases by as little as 0.1. Up to this stage, the transition from $S$+$S$ to $QP$ mode resembles the case $(12,0.9)$ shown in figure~\ref{fig:stability_map} (\textit{b}).  Beyond this point, however, the complex-conjugate pair, instead of moving towards the unit circle, migrates progressively further away from it as $\Rey_m$ increases. At $\Rey_m=100$, a single unstable $QP$ mode is obtained with $\mu_{0,1}=1.116\pm 0.0838i$,  
and the corresponding fully developed DNS shows typical QP characteristics (figure not shown). Perhaps interestingly, further increasing $\Rey_m$ causes the flow to transition back to a coexistence regime of unstable $S$+$S$ mode at $\Rey_m\approx104.1$.

\subsection{Mode transition and interrelation} \label{sec:Discussion}

\begin{figure}
    \centering
    \includegraphics[width=\linewidth]{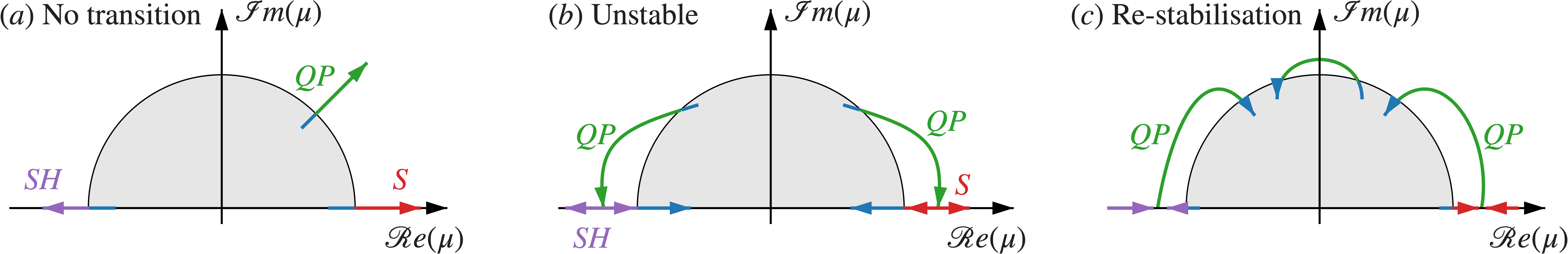}
    \caption{Pathway patterns of the leading Floquet multiplier as $\Rey_m$ increases from a subcritical value to $\Rey_m=100$ for a given $(KC, m)$ condition. ($a$): \emph{No transition}, ($b$): \emph{Unstable transition}, and ($c$) \emph{Re-stabilising transition}. Coloured arrows indicate the classifications of the Floquet mode: synchronous ($S$, red), quasi-periodic ($QP$, green), subharmonic ($SH$, purple), and stable (blue).  
    }
    \label{fig:migration_sketch}
\end{figure}

The possible loci of unstable leading Floquet multipliers can be summarised into three categories, as $\Rey_m$ is gradually increased to $\Rey_m=100$ for a given $(KC,m)$ condition. These are  illustrated schematically in figure~\ref{fig:migration_sketch}.

\emph{No transition:} The base flow bifurcates as a single
mode, with a leading multiplier $\mu_0$, or a complex-conjugate pair ($\mu_0$, $\mu_1$), exiting the unit circle while retaining the same mode, namely $S$, $QP$, or $SH$, figure~\ref{fig:migration_sketch} ($a$). $|\mu_0|$ grows monotonically for $S$ and $QP$, often non-monotonically for $SH$, and no subsequent bifurcation event occurs; that is, no additional multiplier exits the unit circle as a sub-leading unstable mode. This underlies the three primary archetypes of \S\ref{sec:routes}, and explains the formation mechanism of all the single $S$ mode cases (except $(12,0.4)$), all $QP$ mode cases (except for (10,0.6), which has a transient $SH$ mode) 
and $SH$ mode cases for $KC\le7$ and $(9,0.7)$ in figure~\ref{fig:stability_pattern} (\textit{c}). 

\emph{Unstable transition:} This category explains the formation of coexisting unstable modes at $\Rey_m=100$ as discussed in \S\ref{sec:coexist}. The base flow bifurcates becoming $QP$, and subsequently the complex-conjugate pair coalesce on the real axis and split into two real multipliers ($\mu_0$ and $\mu_1$). The unstable $QP$ mode thus transitions to either $S{+}S$ or $SH{+}SH$ mode. Subsequently $|\mu_0|$ increases and $|\mu_1|$ decreases, as they migrate in opposite directions along the same axis, figure~\ref{fig:migration_sketch}($b$). If the sub-leading multiplier $\mu_1$ enters the unit circle and becomes stable at $\Rey_m<100$, a second unstable branch emerges that follows the single $QP$ mode locus depicted in figure~\ref{fig:migration_sketch} (\textit{a}), giving rise to the coexisting unstable mode $S{+}QP$ or $SH{+}QP$. 
Mode $QP{+}QP$ is not observed at $\Rey_m=100$, although it can occur at a lower $\Rey_m$, exemplified by the case $(12,0.3)$ at $\Rey_m=95$ shown in figure~\ref{fig:floquet_compiled_coexist} (\textit{a}). In contrast, the combination of $S{+}SH$ cannot arise from this mechanism: a single complex-conjugate pair colliding on the real axis splits into two multipliers of the same sign and cannot produce one positive and one negative multiplier simultaneously. An $S{+}SH$ combination would instead require two independent crossing events which is not observed in the present parameter range.

\emph{Re-stabilising transition:} This category describes the mechanism by which a post-bifurcated unstable flow re-stabilises as $\Rey_m$ increases, figure~\ref{fig:migration_sketch} ($c$). The leading multiplier $\mu_0$ may re-enter the unit circle either along the locus of a single bifurcated $QP$ mode -- for most cases in the upper-right re-stabilisation sub-region in figure~\ref{fig:stability_map}, the only exception being $(12,0.9)$ which initiates from the $S{+}S$ mode first -- or 
on the negative side of the complex plane from the $SH{+}SH$ state, as observed for the cases residing within the lower-left re-stabilisation sub-region. All three pathways lead to re-stabilisation through the $QP$ mode.

Figure~\ref{fig:stability_map} shows that even at a constant $\Rey_m$, the effect of $U_c$ on the stability of the oscillatory flow is very intricate for $KC\ge7$. As $m$ increases to unity, the unstable mode transitions from $S$ (or mode coexistence at large $KC$), $QP$, $SH$ and to re-stabilisation in general. This pattern defines rather complicated boundaries between neighbouring mode types. Although the unstable $(KC,m)$ conditions located near these boundaries are assigned a definite unstable mode type, they often retain features of neighbouring cases at lower $\Rey_m$, as manifested by the locus of $\mu_0$, suggesting interrelation between modes.   

\begin{figure}
    \centering
    \includegraphics[width=1\linewidth]{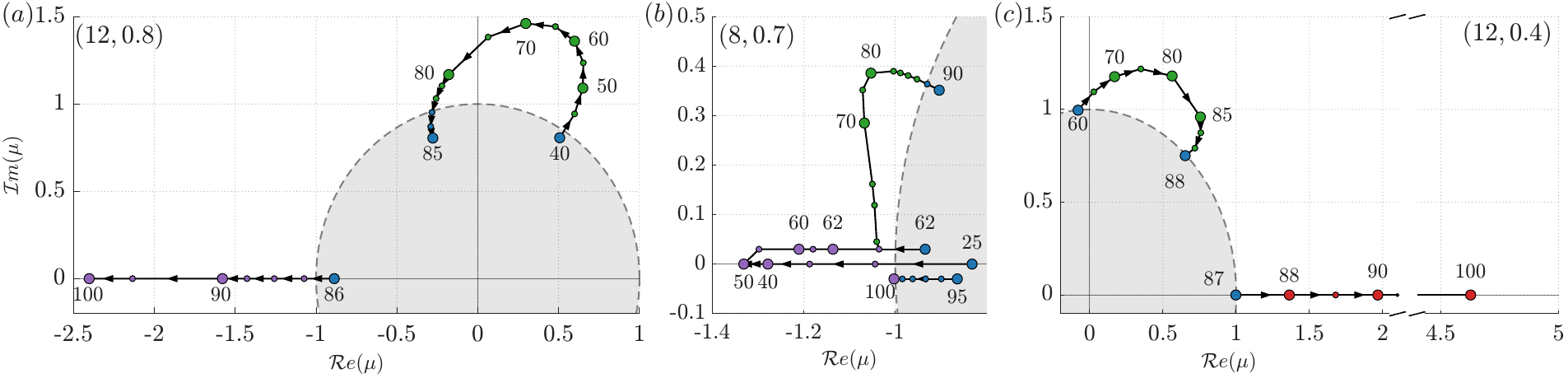}
    \caption{Loci of the leading Floquet multipliers in the complex plane for three cases as $\Rey_m$ increases at given $(KC, m)$ condition. Labels give $\Rey_m$ values; arrows show the direction of increasing $\Rey_m$.  ($a$):$(12, 0.8)$, ($b$):  $(8, 0.7)$,  ($c$): $(12, 0.4)$. In (\textit{b}) two ranges of $\Rey_m$ are shifted away from the negative real axis for clarity. Marker colouring follows the instability classifications in figure~\ref{fig:stability_map} and the dashed curve marks the boundary of the unit circle.
    }
    \label{fig:bound}
\end{figure}

For example, loci of $\mu_0$ for $SH$ cases at $(KC>9,m=0.8)$ demonstrate a history of re-stabilisation at $\Rey_m<100$. A representative of such cases, $(12,0.8)$, close to the upper-right re-stabilisation region, is shown in figure~\ref{fig:bound} (\textit{a}) where stabilisation occurs through the $QP$ mode at $\Rey_m\approx 83$, followed by a second bifurcation to the $SH$ mode at $\Rey_m\approx 87$. 

Located near the boundary with the $QP$ region, $SH$ cases of $(KC>9,m=0.7)$ exhibit a prior $QP$ instability, similar to that illustrated in figure~\ref{fig:stability_pattern} (\textit{d}). The case $(8,0.7)$, figure~\ref{fig:bound} (\textit{b}), is a primarily unstable $SH$ mode, but is at the junction of $QP$ and re-stabilisation regions, and therefore demonstrates features of re-stabilisation following $SH$ and $QP$ stages, akin to figure~\ref{fig:restable_pattern} (\textit{c}). Within $90<\Rey_m<100$, flow remains re-stabilised, until at $\Rey_m\approx100$, the flow bifurcates for a second time becoming $SH$.  Another example is case $(12,0.4)$, which shows large $|\mu_0|$ (bigger size marker) single $S$ unstable mode, but is located in the coexistence region (instead of the synchronous region) surrounded by $QP$ and $S{+}QP$ modes. The locus of $\mu_0$ is presented in figure~\ref{fig:bound} (\textit{c}), demonstrating the characteristics of both. Note this case did not go to re-stabilisation since just before the complex-conjugate pair enters the unit circle at $\Rey_m\approx88$, another multiplier exits along the positive real axis at $\Rey_m\approx87$.   

\begin{figure}
    \centering
    \includegraphics[width=0.7\textwidth]{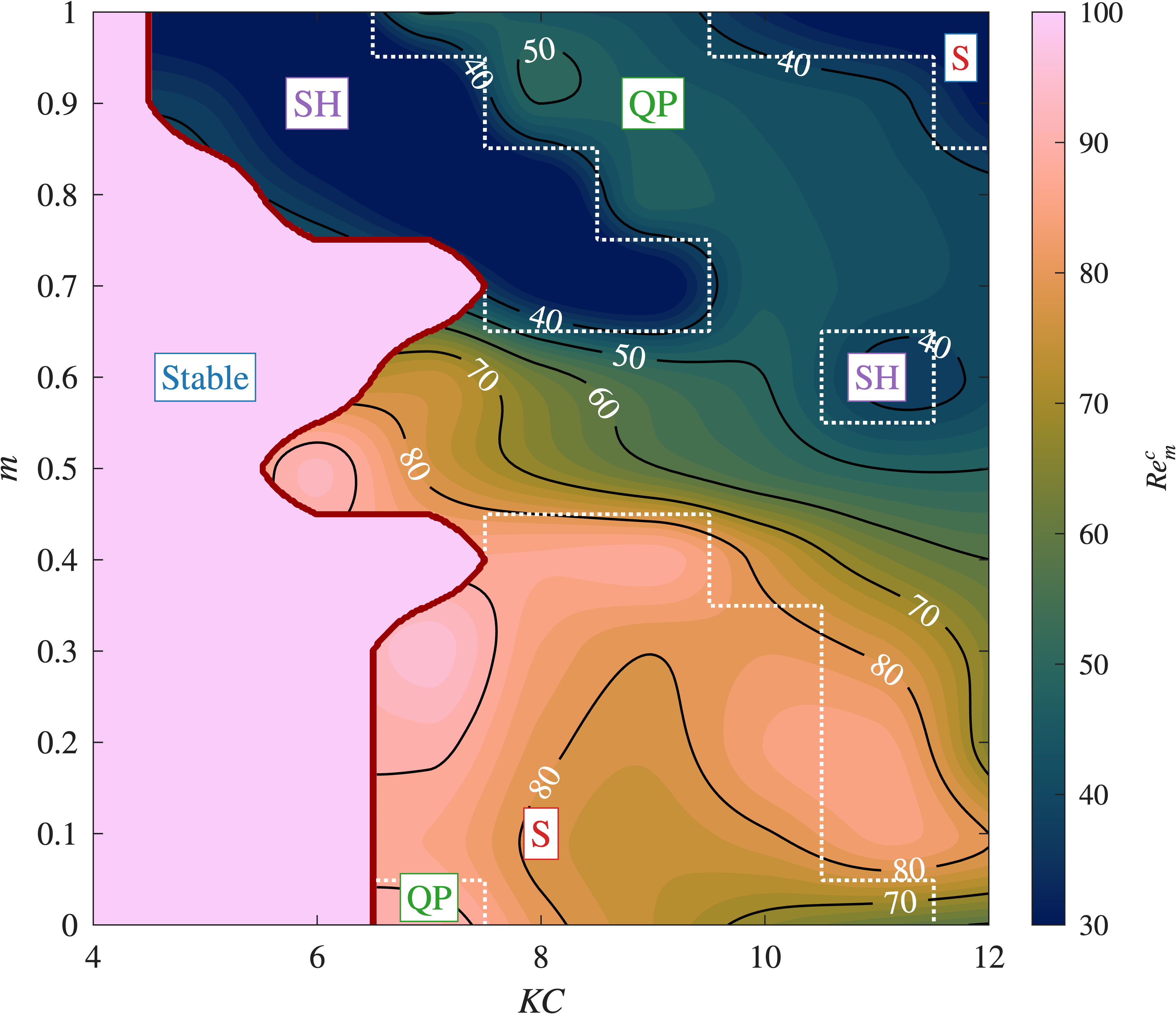}
    \caption{Estimated dependence of critical Reynolds number $Re_m^c$ on $(KC,m)$. The field is interpolated onto a fine mesh using the modified Akima piecewise cubic Hermite scheme \citep{Akima1970}. Filled contours and thin black lines show $Re_m^c$ value. The cases in the stable region on the left of the heavy dark-red neutral stability curve do not bifurcate for $Re_m\le100$. Dotted white lines crudely delineate the bifurcation mode boundaries at instability onset. 
    }
    \label{fig:Rec_type} 
  \end{figure}

In the present study, no attempt is made to determine the exact primary critical Reynolds number, $\Rey_m^c$, at which the leading multiplier $\mu_0$ first exits the unit circle for each $(KC,m)$ condition. Nevertheless, an estimate of $\Rey_m^c$ is presented in figure~\ref{fig:Rec_type}, with a resolution of $\Delta \Rey_m^c=\pm2.5$, together with the corresponding bifurcation mode, which
is different from that at $\Rey_m=100$. The distribution of $\Rey_m^c$ is obtained by interpolating the grid data onto a finer mesh using the modified Akima piecewise cubic Hermite scheme \citep{Akima1970}. An apparent discontinuity in $\Rey_m^c$ from $\Rey_m^c>100$ to $\Rey_m^c<50$ is observed across the boundary in the region $KC\le7$ and $m\ge0.7$, suggesting that the current parameter resolution, $\Delta KC=1$ and $\Delta m=0.1$, is insufficient to smoothly resolve the variation in $\Rey_m^c$ there. 
All bifurcations identified in the present study are codimension-1 indeed, with a single multiplier crossing the unit circle, consistent with the single-mode  partition shown in the figure \ref{fig:Rec_type}.

It is important to reiterate that the present study is conducted strictly from a 2D perspective. As noted in \S\ref{sec:methods}, 3D secondary instabilities are, for example, known to destabilise the 2D limit cycle at $\Rey \approx 190$ for steady current \citep{Barkley1996,Blackburn2003}.
In regions of parameter space where $\Rey_p$ approaches or exceeds this threshold, the 2D Floquet analysis presented here may therefore be pre-empted by the onset of 3D modes. This caveat applies most directly to the high $\Rey_p$ conditions including the re-stabilisation flow regime. For example, although it is reasonable to speculate that the re-stabilised cases observed at $\Rey_m=100$ may undergo a second bifurcation at higher $\Rey_m$, akin to the mechanism in figure~\ref{fig:bound} (\textit{a}--\textit{b}), the onset of 3D instabilities could occur before such a 2D transition is reached.

\section{Concluding remarks} \label{sec:conclusion}

The effect of a superposed collinear aligned current on the stability of an oscillatory flow past a circular cylinder is studied from a 2D standpoint 
with the global stability state 
of this combined flow when $Re_m=100$ 
revealed using 
linear Floquet theory in tandem with 
the fully nonlinear behaviour of the linearly unstable flow evaluated via DNS. The 
loci of the leading Floquet multipliers is investigated by progressively reducing $Re_m$ from 100 to subcritical. 

A key finding is that the superposition of a steady current enriches the symmetry-breaking pathway. When the ratio of steady to oscillatory velocity component exceeds $m>0.5$, a subharmonic route (mode $SH$) becomes accessible: the leading Floquet multiplier exits the unit circle at $\mu_0=-1$, giving rise to a period-doubled unstable mode that is otherwise theoretically forbidden in the purely oscillatory flow case. This mode can emerge directly from the primary bifurcation, as in the case $(KC,m)=(9,0.7)$, or arise through a transition from a $QP$ mode with increasing $Re_m$, as observed for the case $(10,0.7)$. An interesting characteristic of the $SH$ mode is that, beyond bifurcation, the trajectory of $\mu_0$ along the negative real axis is often non-monotonic. Consequently, the instability growth at $Re_m=100$ can be weaker than that at lower values of $Re_m$, implying that the flow may destabilise more slowly despite the higher Reynolds number.

The stability landscape in $(KC,m)$ parameter space whenn $Re_m=100$, shown in figure~\ref{fig:stability_map}, reveals three notable features. First, flows with $KC\le6$ are intrinsically stable for all $m<1$. For $KC>6$, three primary unstable modes are identified, namely $S$, $QP$, and $SH$, which populate the stability landscape in approximate succession as $m$ increases from $0$ to $0.8$. Second, two sub-regions of re-stabilisation are embedded within the general stable regime. These flows, occurring typically for $m>0.7$, follow a sequence in which $\mu_0$ first exits and subsequently re-enters the unit circle as $\Rey_m$ increases. As a consequence, the flow recovers a $Z_2$-symmetric state at peak Reynolds numbers approaching $190$, despite the fact that the corresponding steady-current and purely oscillatory wakes are each individually unconditionally unstable in two dimensions. This mechanism gives rise to the highly non-trivial shape of the neutral stability boundary when $Re_m=100$. Third, a region of simultaneous mode coexistence is identified for $KC\ge10$ and $0.1\le m\le0.4$, within which the dominant combinations are $S+S$ and $S+QP$.

The value $Re_m=100$, for which figure~\ref{fig:stability_map} is constructed, lies beyond the critical Floquet threshold and is, in some cases, substantially larger than the critical value, which can be as low as $Re_m<30$; see figure~\ref{fig:stability_pattern} (\textit{c}). Nevertheless, for cases with a single unstable mode, the identified linear mode generally predicts the fully developed nonlinear DNS results well; compare figure~\ref{fig:stability_pattern} (\textit{d}) with figure~\ref{fig:ts_kc10_m0p7_sub_vorticity}. For cases with mode coexistence, the DNS results suggest that the flow preferentially evolves toward the $QP$ manifestation in the nonlinearly saturated state. This tendency persists for $S+QP$ cases where $|\mu_0(S)| > |\mu_0(QP)|$, such as $(12,0.3)$, and even for $S+S$ cases in the absence of a $QP$ mode (at $Re_m=100$), such as $(11,0.1)$; see figure~\ref{fig:ts_kc12_m0p3_kc11_m0p1_co}. 

\begin{bmhead}[Acknowledgments]
The authors are grateful to Professor Dwight Barkley for the interest he has shown in this work, and wish to thank him for his helpful and insightful suggestions. Use was made of the Hamilton HPC Service at Durham University.
\end{bmhead}

\begin{bmhead}[Funding]
The work was conducted under the Aura CDT programme, funded by EPSRC and NERC, grant number EP/S023763/1 and project reference 2881662.
\end{bmhead}

\begin{bmhead}[Declaration of interests]
The authors report no conflict of interest.
\end{bmhead}
\begin{bmhead}[Author ORCIDs]

Geng Chen  https://orcid.org/0000-0003-4863-9199; 

Lian Gan  https://orcid.org/0000-0002-4948-4523;

Philip H. Gaskell https://orcid.org/0000-0001-8572-8997 .
\end{bmhead}

\appendix
\begin{appen}
\counterwithin{figure}{section}
\counterwithin{table}{section}
\section{Validation of the computational framework}\label{appA}
The computational framework described in \S\ref{sec:methods}  has been carefully validated in terms of force coefficients and velocity profiles for a purely oscillatory flow, and force coefficients and shedding frequency for a steady current.

\subsection{Oscillatory flow}\label{sec:oscValid}
Mesh convergence was assessed for the flow case $(KC, m) = (5,0)$ when $\Rey_m=100$ by varying the domain radius $R_d / D \in \{70, 80, 100\}$, the number of circumferential mesh elements around the cylinder $N_c \in \{24, 36\}$, and the polynomial order $N_p \in \{6, 7, 8\}$ for the spectral/\textit{hp} element method. The first radial element thickness at the cylinder surface was fixed at $0.03D$, similar to \citet{Ren2019}. 

Making use of the two-term Morison equation \citep{Chen2026}:
\begin{equation}
\label{eq:relativeM}
    F_x = \frac{1}{2}\mathrm{C}_{\mathrm{D}} \rho D \left(U_m  \right)|U_m|+ \rho \frac{\pi \mathrm{D}^2}{4} \mathrm{C}_{\mathrm{M}}  \dot{U}_m,
\end{equation}
where $\dot{U}_m = 2\pi (U_m/T)\cos(2\pi t/T)$, according to (\ref{eqn:BCU}), the key convergence performance indicators are the fitted drag ($C_D$) and inertia ($C_M$) coefficients, which are extracted by a least-squares fitting of $F_x$ obtained from DNS with (\ref{eq:relativeM}) over $t>200T$. 

Table~\ref{tab:cdcm_validation} compares the values of $C_D$ and $C_M$ obtained using different domain sizes with the reference results of \citet{DUTSCH1998}, \citet{Zhao2014}, and \citet{Tong2015}. Very good agreement is achieved across all three domain configurations. Variations in $N_c$ and $N_p$ lead to changes of less than $1\%$ in both $C_D$ and $C_M$, and are thus not shown. On this basis, the solution domain utilised throughout the present study had the characteristics: $R_d = 70D$, $N_c = 24$, and $N_p = 6$.

\begin{table}
  \begin{center}
\def~{\hphantom{0}}
  \begin{tabular}{lcccc}
      
      Case & Domain size& $C_D$  & $C_M$ \\ \hline
      Present           & $R_{d} = 70D$            & 2.049  & 2.449 \\
      Present           & $R_{d} = 80D$            & 2.049  & 2.448 \\
      Present          & $R_{d} = 100D$           & 2.034  & 2.451 \\
      \citet{DUTSCH1998} & --              & 2.030  & 2.460 \\
      \citet{Zhao2014}   & 30D $\times$ 30D & 2.040 & 2.480 \\
      \citet{Tong2015} & 100D $\times$ 100D & 2.110 & 2.420 \\ 
  \end{tabular}
  \caption{Comparison of $C_D$ and $C_M$ for a single cylinder in an oscillatory flow at $(KC,m) = (5,0)$, $\Rey_m=100$ for all studies.}
  \label{tab:cdcm_validation}
  \end{center}
\end{table}

Figure~\ref{fig:cdcm35} compares the computed values of $C_D$ and $C_M$ at $\beta = 35$ with the experimental measurements of \citet{Kuhtz1996} and the numerical results of \citet{DUTSCH1998} and \citet{Tong2015} over the range of $KC$ considered in the present study, demonstrating good quantitive agreement. The velocity profiles at four streamwise locations for $(KC,\beta)=(5,20)$, $\Rey_m=100$, against the experimental measurements reported by \citet{DUTSCH1998} is also compared  in \citet{Chen2026}. 
In addition, the computed fully developed streakline patterns (not shown) agree closely with the regime classifications of \citet{Tatsuno1990}, further confirming the accuracy of the present solver for purely oscillatory flow.

\begin{figure}
    \centering
    \includegraphics[width=1\linewidth]{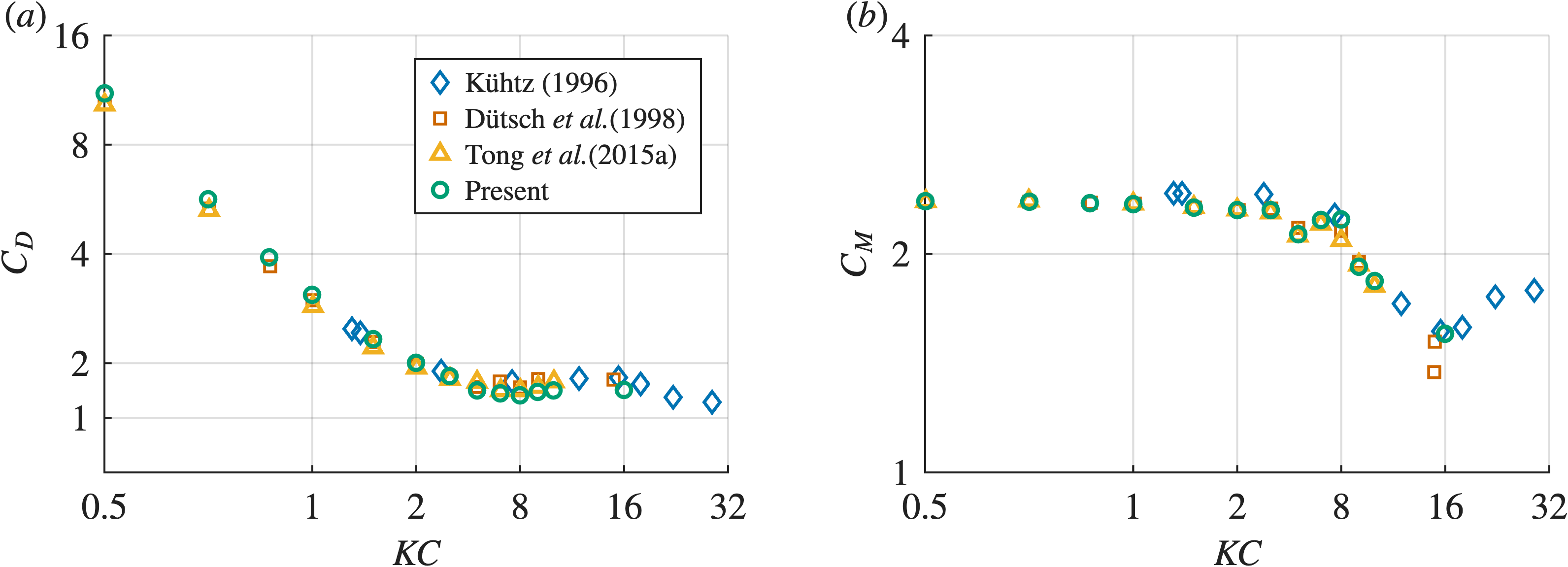}
    \caption{Comparison of $C_D$ (\textit{a}) and $C_M$ (\textit{b}) for $\beta =35$, with \citet{Kuhtz1996}, \citet{DUTSCH1998} and \citet{Tong2015}. } 
    \label{fig:cdcm35}
\end{figure}

\subsection{Steady current} \label{sec:steadyValid}
Table~\ref{tab:cdclst_validation} compares the computed mean value of $C_d$, the root-mean-square of $C_l$ (denoted as $C'_l$); see equation \ref{eqn:cdcl},  
and the non-dimensional vortex-shedding frequency, $St = fD/U_c$ 
for a steady current ($m = 0$) past a circular cylinder when $\Rey = 100$ (based on $U_c$) with values reported in the literature. The present results agree with the experimental and numerical studies of \citet{Williamson1989}, \citet{Li2009}, and \citet{Chen2020} to within $1\%$ for $C_d$ and $St$, and to within $2\%$ for $C'_l$.
\begin{table}
  \begin{center}
\def~{\hphantom{0}}
  \begin{tabular}{lccc} 
      Case                & $C_d$  & $C'_l$  & $St$  \\ \hline
      \citet{Williamson1989}  & --     & --     & 0.164 \\
      \citet{Li2009}          & 1.336  & --     & 0.164 \\
      \citet{Chen2020}        & 1.337  & 0.230  & 0.163 \\
      Present              & 1.323  & 0.226  & 0.165 \\ 
  \end{tabular}
  \caption{Comparison of $C_d$, $C'_l$, and $St$ for a steady current ($m=0$) past a circular cylinder when $Re = 100$.}
  \label{tab:cdclst_validation}
  \end{center}
\end{table}

\section{Validation of the Floquet stability framework}\label{sec:FloquetValid}
\counterwithin{figure}{section}

The Floquet solver described in \S\ref{sec:floquetM} is validated against the critical Floquet mode for pure oscillatory flow ($m=0$) reported by \citet{Elston2006}. Figure~\ref{fig:oscfloquetvalid} compares the location of the marginal leading Floquet multiplier, $|\mu_0|=1$, together with the corresponding phase, $\theta=\arg(\mu_0)$, against the neutral stability curves identified by \citet{Elston2006} in $(\beta,KC)$ parameter space. Excellent agreement is obtained for both quantities, with discrepancies remaining within plotting accuracy throughout the range considered.

\begin{figure}
    \centerline{\includegraphics[width=0.9\linewidth]{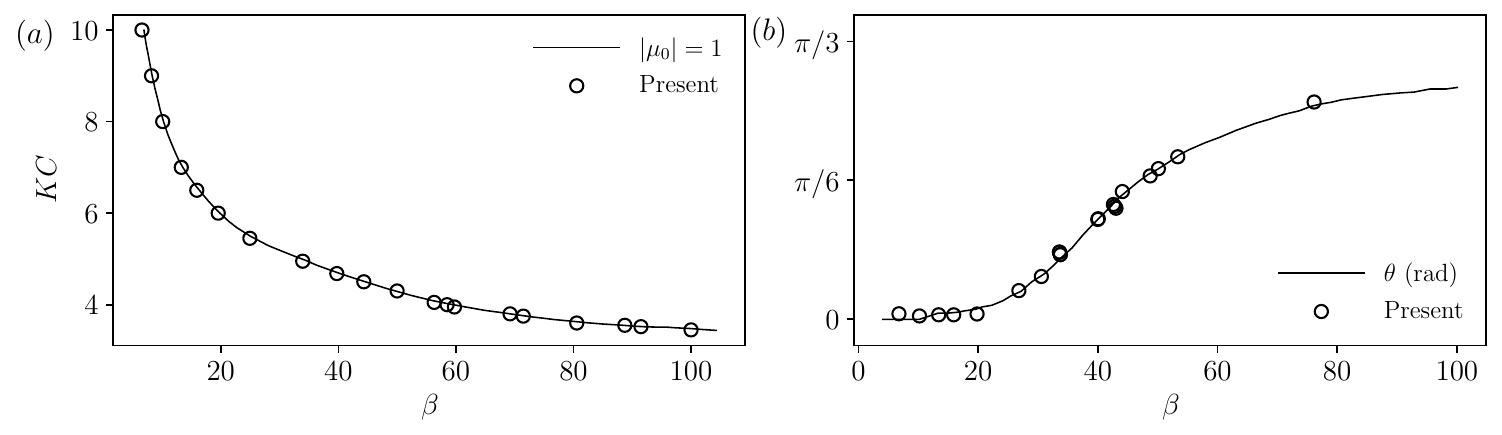}}
        \caption{Marginal leading Floquet multiplier $|\mu_0|=1$ and the corresponding phase $\theta=\arg(\mu_0)$ for purely oscillatory flow ($m=0$) in $(\beta, KC)$ parameter space. Solid lines denote the neutral stability curves identified by \citet{Elston2006}; symbols represent the present results.}
    \label{fig:oscfloquetvalid}
\end{figure}

Figure~\ref{fig:oscfloquetMvalid} further compares the spatial structures of the critical Floquet eigenfunctions for two representative cases: a $S$ mode at $(\beta,KC)=(13.75,7)$ and a $QP$ mode at $(40,4.7)$. The computed eigenfunction topologies exhibit excellent agreement with those reported by \citet{Elston2006}, thereby providing further validation of the Floquet framework employed in the work reported here.

\begin{figure}
    \centerline{\includegraphics[width=0.75\linewidth]{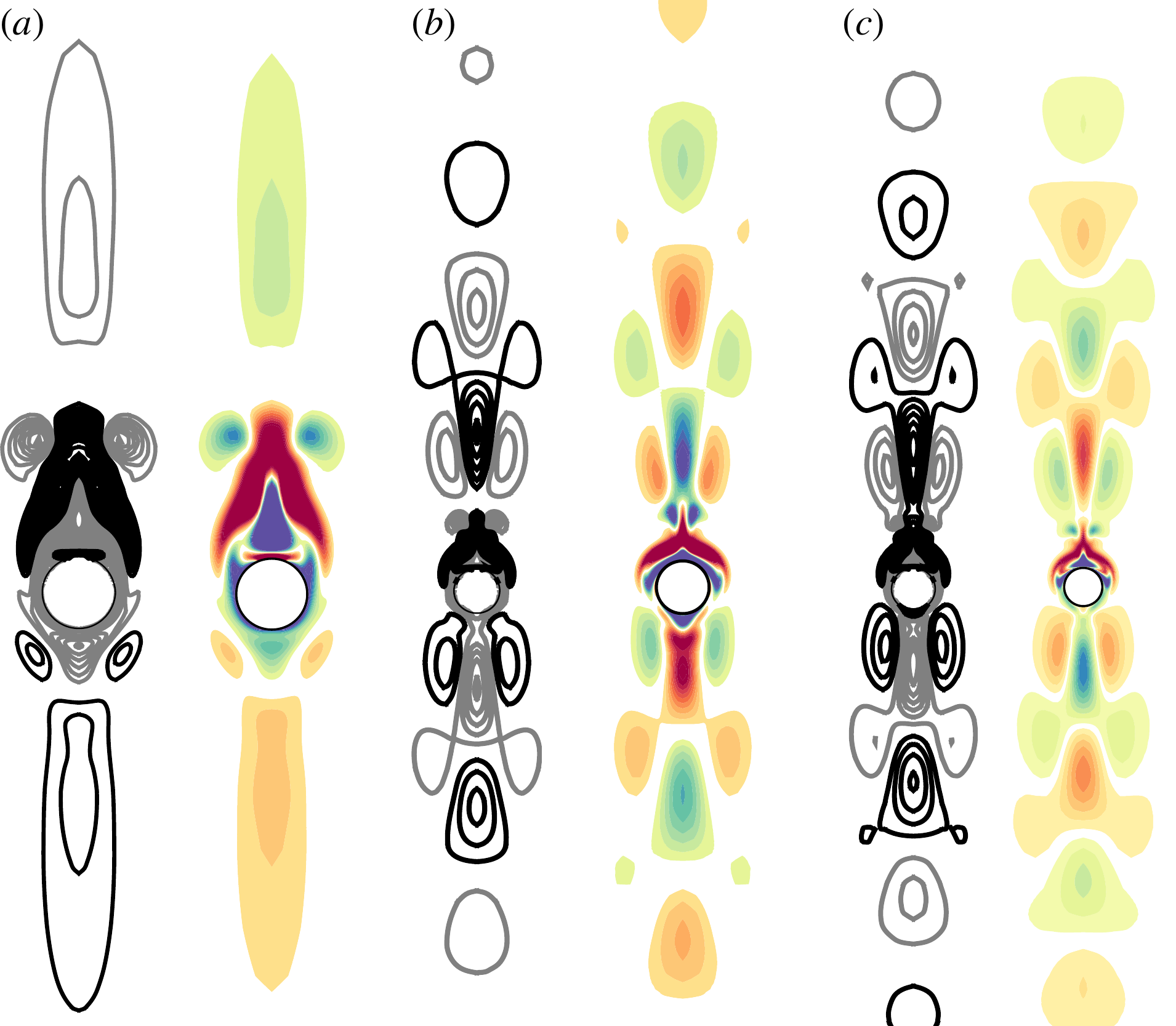}}
        \caption{Comparison of critical Floquet eigenfunction for purely oscillatory flow past a circular cylinder. ($a$): $\mathcal{R}e[\tilde{\omega}_z]$, where $\tilde{\omega}_z=\nabla\times\tilde{\mathbf{u}}$, for the $S$ mode $(\beta, KC) = (13.75, 7)$; ($b$) and ($c$) $\mathcal{R}e[\tilde{\omega}_z]$ and $\mathcal{I}m[\tilde{\omega}_z]$, respectively, for the $QP$ mode $(\beta, KC) = (40, 4.7)$. In each panel, the plot on the left is the result of \citet{Elston2006}, while that on the right is the corresponding one from the present study.}
    \label{fig:oscfloquetMvalid}
\end{figure}

\end{appen}\clearpage

\bibliographystyle{jfm}
\bibliography{jfm}

\end{document}